\documentclass[]{spie}  
\usepackage[pdftex]{graphicx}
\usepackage{epstopdf}
\usepackage{amsmath}
\usepackage{pstricks}
\usepackage{lscape}
\usepackage[T1]{fontenc}
\usepackage{ae,aecompl}
\usepackage{remreset}
\usepackage{url}
\usepackage{wasysym}

\title{Feedhorn-Coupled TES Polarimeter Camera Modules at 150 GHz for CMB Polarization Measurements with SPTpol} 

\author{
J.W.~Henning\supit{c}, P.~Ade\supit{b}, K.A.~Aird\supit{r}, J.E.~Austermann\supit{c}, J.A.~Beall\supit{d}, D.~Becker\supit{d}, B.A.~Benson\supit{a,s}, L.E.~Bleem\supit{a,q}, J.~Britton\supit{d}, J.E.~Carlstrom\supit{a,e,q,s,t}, C.L.~Chang\supit{a,e,s}, H-M.~Cho\supit{d}, T.M.~Crawford\supit{a,t}, A.T.~Crites\supit{a,t}, A.~Datesman\supit{g}, T.~de Haan\supit{f}, M.A.~Dobbs\supit{f}, W.~Everett\supit{a}, A.~Ewall-Wice\supit{a,q}, E.M.~George\supit{h}, N.W.~Halverson\supit{c,p}, N.~Harrington\supit{h}, G.C.~Hilton\supit{d}, W.L.~Holzapfel\supit{h}, J.~Hubmayr\supit{d}, K.D.~Irwin\supit{d}, M.~Karfunkle\supit{a,q}, R.~Keisler\supit{a,q,s}, J.~Kennedy\supit{f}, A.T.~Lee\supit{h}, E.~Leitch\supit{a}, D.~Li\supit{d}, M.~Lueker\supit{j}, D.P.~Marrone\supit{o}, J.J.~McMahon\supit{k}, J.~Mehl\supit{a,s}, S.S.~Meyer\supit{a,q,s,t}, J.~Montgomery\supit{a,q}, T.E.~Montroy\supit{m}, J.~Nagy\supit{m}, T.~Natoli\supit{a,q}, J.P.~Nibarger\supit{d}, M.D.~Niemack\supit{d}, V.~Novosad\supit{g}, S.~Padin\supit{a}, C.~Pryke\supit{l}, C.L.~Reichardt\supit{h}, J.E.~Ruhl\supit{m}, B.R.~Saliwanchik\supit{m}, J.T.~Sayre\supit{m}, K.K.~Schaffer\supit{n}, E.~Shirokoff\supit{j}, K.~Story\supit{a,q}, C.~Tucker\supit{b}, K.~Vanderlinde\supit{f}, J.D.~Vieira\supit{j}, G.~Wang\supit{e}, R.~Williamson\supit{a,s}, V.~Yefremenko\supit{e,g}, K.~W.~Yoon\supit{d}, E.~Young\supit{h}
\skiplinehalf
\supit{a} Kavli Institute for Cosmological Physics, Department of Physics, Enrico Fermi Institute, The University of Chicago, Chicago, IL 60637, USA
\skiplinehalf
\supit{b} Cardiff School of Physics and Astronomy, Cardiff University, Cardiff, United Kingdom
\skiplinehalf
\supit{c} Department of Astrophysical and Planetary Sciences, University of Colorado, Boulder, CO 80309, USA
\skiplinehalf
\supit{d} NIST, Boulder, CO 80305, USA
\skiplinehalf
\supit{e} High Energy Physics Division, Argonne National Laboratory, Argonne, IL 60439, USA
\skiplinehalf
\supit{f} McGill University, Montreal, Quebec, Canada
\skiplinehalf
\supit{g} Materials Science Division, Argonne National Laboratory, Argonne, IL 60439, USA
\skiplinehalf
\supit{h} University of California, Berkeley, 151 LeConte Hall Berkeley, CA 94720, USA
\skiplinehalf
\supit{j} California Institute of Technology, Pasadena, CA 91125, USA
\skiplinehalf
\supit{k} University of Michigan, Ann Arbor, MI, USA
\skiplinehalf
\supit{l} University of Minnesota, Minneapolis, MN 55455, USA
\skiplinehalf
\supit{m} Case Western Reserve University, Cleveland, OH 44106, USA
\skiplinehalf
\supit{n} School of the Art Institute of Chicago, Chicago, IL 60603, USA
\skiplinehalf
\supit{o} Steward Observatory, University of Arizona, 933 North Cherry Avenue, Tucson, AZ 85721, USA
\skiplinehalf
\supit{p} Department of Physics, University of Colorado, Boulder, CO 80309, USA
\skiplinehalf
\supit{q} Department of Physics, University of Chicago, 5640 South Ellis Avenue, Chicago, IL 60637, USA
\skiplinehalf
\supit{r} University of Chicago, 5640 South Ellis Avenue, Chicago, IL 60637, USA
\skiplinehalf
\supit{s} Enrico Fermi Institute, University of Chicago, 5640 South Ellis Avenue, Chicago, IL 60637, USA
\skiplinehalf
\supit{t} Department of Astronomy and Astrophysics, University of Chicago, 5640 South Ellis Avenue, Chicago, IL 60637, USA
}

\authorinfo{Send correspondence to J.W. Henning: E-mail: Jason.Henning@colorado.edu}

\begin{document} 
\maketitle 

\begin{abstract}
The SPTpol camera is a dichroic polarimetric receiver at 90 and 150~GHz.  Deployed in January 2012 on the South Pole Telescope (SPT), SPTpol is looking for faint polarization signals in the Cosmic Microwave Background (CMB).  The camera consists of 180 individual Transition Edge Sensor (TES) polarimeters at 90~GHz and seven 84-polarimeter camera modules (a total of 588 polarimeters) at 150~GHz.  We present the design, dark characterization, and in-lab optical properties of the 150~GHz camera modules.  The modules consist of photolithographed arrays of TES polarimeters coupled to silicon platelet arrays of corrugated feedhorns, both of which are fabricated at NIST-Boulder.  In addition to mounting hardware and RF shielding, each module also contains a set of passive readout electronics for digital frequency-domain multiplexing. A single module, therefore, is fully functional as a miniature focal plane and can be tested independently.  Across the modules tested before deployment, the detectors average a critical temperature of 478~mK, normal resistance $R_N$ of 1.2~$\Omega$, unloaded saturation power of 22.5~pW, (detector-only) optical efficiency of $\sim$ 90\%, and have electrothermal time constants < 1 ms in transition.
\end{abstract}



\section{INTRODUCTION}\label{intro} 
Temperature anisotropies in the Cosmic Microwave Background (CMB) have been measured with high precision from large angular scales \cite{Chiang_2010, Larson_2011} down to scales of a few arcminutes\cite{Lueker_2010, Das_2011, Keisler_2011} and ongoing experiments continue to push to higher sensitivities.  Measurements of these temperature fluctuations have yielded constraints on bulk properties of the Universe, including its curvature, the matter-energy content, and the spectral distribution of initial matter fluctuations.  The CMB is also weakly polarized, and studying polarization anisotropies can tighten constraints placed on cosmological parameters.  E-mode polarization, divergence-like patterns, was first observed by the DASI experiment in 2002 \cite{Kovac_2002}, and since by several other experiments \cite{Barkats_2005b, Montroy_2006, Bischoff_2008, Pryke_2009, Chiang_2010, Larson_2011}.  B-mode polarization, curl-like patterns, has yet to be detected.

The measured E-mode power spectrum is two orders of magnitude fainter than the temperature anisotropy power spectrum, and the B-mode spectrum is predicted to be fainter by another two orders of magnitude.  With background-limited detectors, ground-based and balloon-borne experiments have moved to ever larger focal planes while simultaneously gaining tighter control on systematics to achieve the requisite sensitivities to detect CMB polarization.  The SPTpol camera, a dichroic polarimetric receiver measuring the CMB in bands centered at 90 and 150 GHz, recently deployed at the South Pole.  With 1536 transition edge sensor (TES) detectors, (768 polarization-sensitive pixels), SPTpol will detect polarization anisotropies down to $\sim 10^{-2} \mu$K$^2$ after the full survey depth is achieved\cite{Bleem_2012}.  This sensitivity will yield constraints on the sum of the neutrino masses $\Sigma m_{\nu}$ and the energy scale of inflation through the tensor to scalar fluctuation ratio $r$\cite{Austermann_2012}.

In this paper we discuss the design and detector properties of the 150 GHz portion of the SPTpol focal plane, which has been split into seven 84-pixel modules.  Each module is an independent camera including coupling feedhorns, radio frequency (RF) shielding, and passive readout electronics.  The paper is organized as follows: Section \ref{Design} describes the module design in detail.  Notable features of both individual hardware elements and the modules as a whole are discussed.  Section \ref{Dark} provides a summary of detector dark properties, including critical temperatures, normal resistances, and saturation powers.  Section \ref{Optical} reviews the lab-measured optical properties of the modules, namely detector optical efficiencies and electrothermal time constants, as well as early optical results for silicon platelet corrugated feedhorn arrays developed at NIST-Boulder.  Finally we summarize the design and module properties in Section \ref{Conclusions}.

\section{Module Design}\label{Design}
\begin{figure}[h]
\begin{center}
\resizebox{0.4\textwidth}{!}{
\includegraphics[width=1\textwidth]{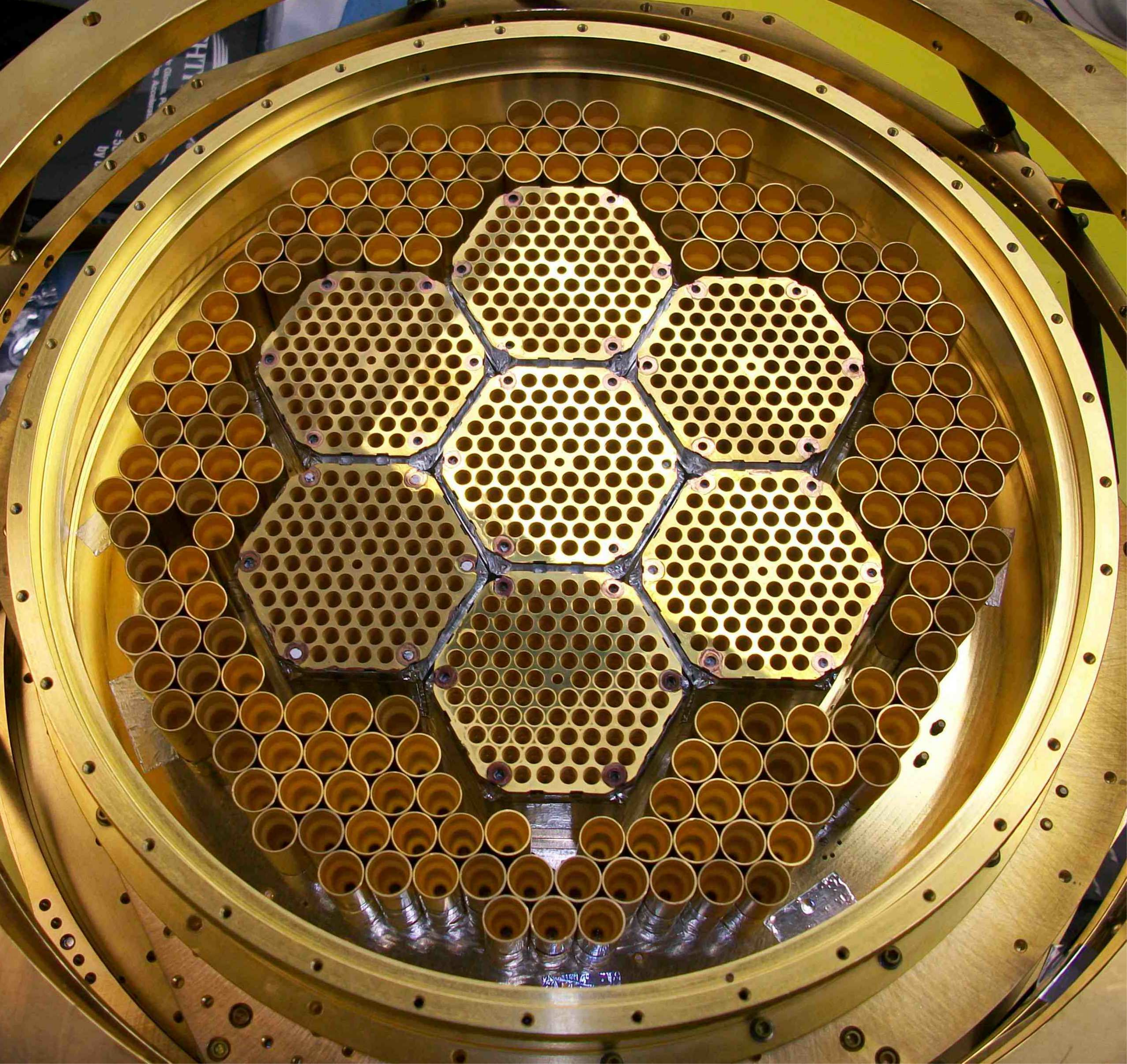}}
\caption{The SPTpol focal plane.  Seven 84-pixel modules of 150 GHz detectors sit at the center of the camera, while 180 individually packaged 90 GHz pixels surround the modules, for a total of 768 polarization-sensitive pixels (1536 detectors).  The focal plane is $\sim$ 225 mm in diameter.}\label{focal_plane}
\end{center}
\end{figure}

A large focal plane poses significant challenges for fabrication, assembly, and testing.  To mitigate these issues, portions of SPTpol were designed with a modular format.  While the 90 GHz band of SPTpol is comprised of 180 polarization-sensitive pixels (360 detectors fabricated at Argonne National Labs) individually packaged with corresponding light-coupling feedhorns \cite{Sayre_2012}, the 150 GHz portion of the focal plane is split into seven identical modules, each containing 2.3 inch wide monolithic feedhorn and detector arrays fabricated at NIST in Boulder, CO.  Figure \ref{focal_plane} shows a front-side view of the SPTpol focal plane where the 150 GHz modules and 90 GHz pixel assemblies are clearly distinguishable.  Modularity has several benefits.  First, maintaining detector uniformity is easier when fabricating across smaller wafers.  Second, a modular design makes possible testing whole units of the focal plane without requiring the full experiment apparatus.  This greatly increases testing throughput and accelerates feedback into the design process.  Finally, the modular units are simple to install into and remove from the SPTpol focal plane, which makes in-field modifications more tractable and timely.  In the following subsections we describe in detail distinguishing characteristics of the module components.

\subsection{Corrugated Silicon Platelet Feedhorn Array}\label{Feeds}
When coupling free space to detectors, corrugated feedhorns exhibit several appealing characteristics, namely high transmission efficiency, low cross-polarization and sidelobes, highly symmetric beam shapes, and wide bandwidths\cite{Clarricoats_1984}.  Since measuring CMB polarization anisotropies requires tight control of systematics, many past experiments with relatively few pixels used individual corrugated feedhorns \cite{Barnes_2002, Kovac_2002, Padin_2002, Jones_2003, Barkats_2005, Bock_2009, Hinderks_2009, Takahashi_2010}.  Modern experiments contain many hundreds of tightly packed pixels, however, and the production of a corresponding monolithic feedhorn array with standard electroforming techniques would be prohibitively difficult and expensive.  Instead, the SPTpol 150 GHz modules contain monolithic arrays of corrugated feedhorns built up from 33 silicon platelets, each 500 $\mu$m thick, which have been stacked and gold-plated.  The arrays were developed and fabricated at NIST-Boulder\cite{Britton_2010, Hubmayr_2012a}.  In addition to the attractive properties that corrugated feedhorns exhibit, these silicon platelet arrays are coefficient of thermal expansion (CTE) matched to the detector arrays (also fabricated on silicon wafers), have lower thermal mass compared to conventional aluminum feedhorns, and maintain high thermal conductivity despite being silicon in bulk due to the gold-plating.

\begin{figure}[t]
\begin{center}
\resizebox{0.85\textwidth}{!}{
\includegraphics[width=1.0\textwidth]{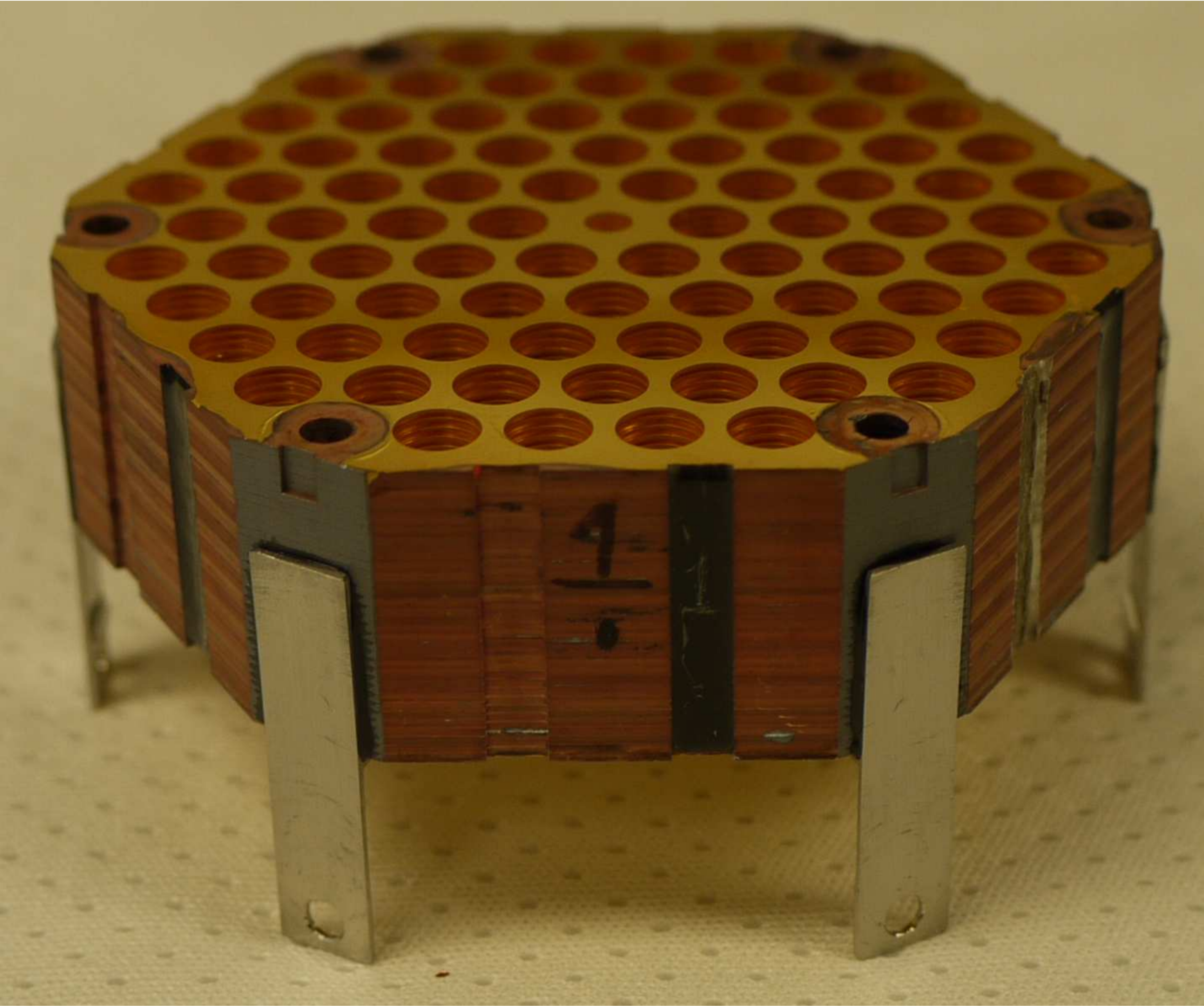}
\includegraphics[width=1.5\textwidth]{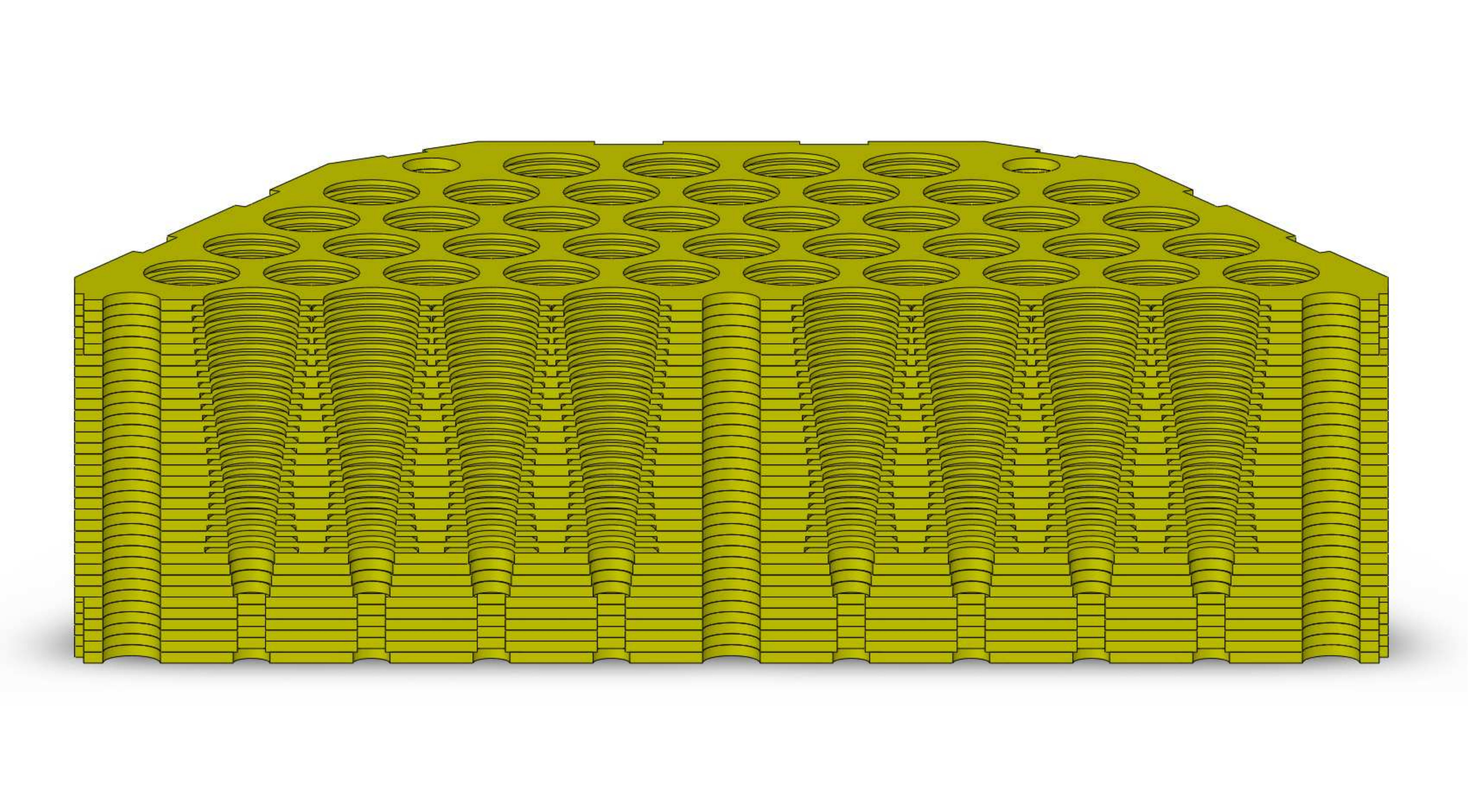}}
\caption{(Left) One of seven 150 GHz corrugated silicon platelet feedhorn arrays in the SPTpol focal plane.  The array is 2.3 inches wide and contains 84 feedhorns with 4.26 mm apertures.  (Right) Cross-sectional view of the feedhorn profiles showing the corrugations and waveguides.}\label{feed_cross-section}
\end{center}
\end{figure}

Each of the seven SPTpol feedhorn arrays are 2.3 inches wide and 16.5 mm tall, contain 84 single-moded corrugated feedhorns with 4.26 mm apertures, and are optimized for a bandpass centered at 145 GHz.  The feedhorns taper off to a section of 1.22 mm wide square waveguide, which is used to define the lower edge of the bandpass, and ends with a single 1.6 mm diameter circular waveguide platelet.  The left of Figure \ref{feed_cross-section} is a picture of one SPTpol feedhorn array, while the right shows a cross-section of a feedhorn array to highlight the feedhorn profile.  300 K vector network analyzer (VNA) measurements reveal that the feedhorns have excellent uniformity in transmission properties, both across a single array and between the seven arrays.  See Section \ref{Optical} below for details.

\subsection{Detector Arrays}\label{Detectors}
\begin{figure}[t]
\begin{center}
\resizebox{0.8\textwidth}{!}{
\includegraphics[width=1.0\textwidth]{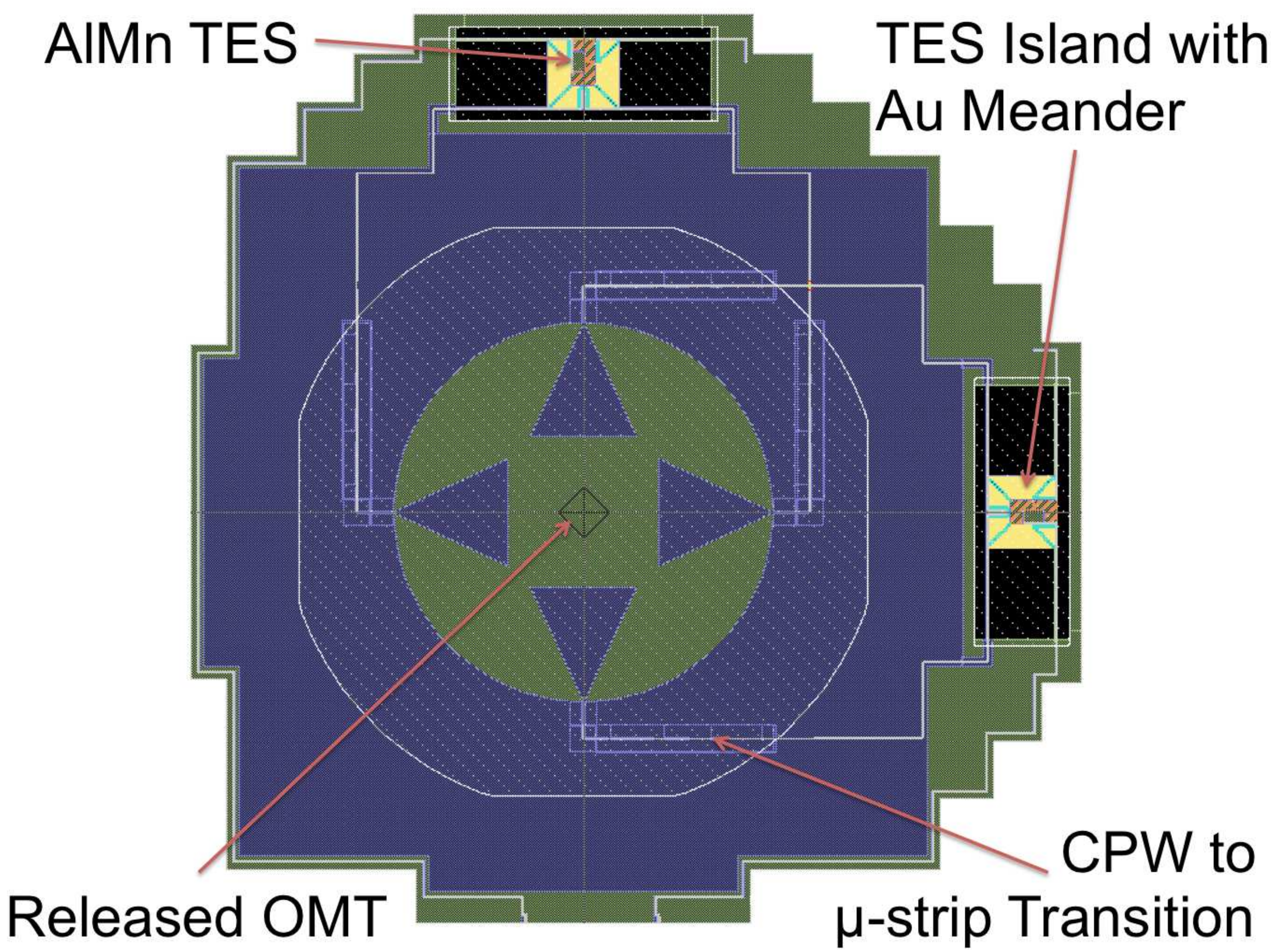}
\includegraphics[width=1.0\textwidth]{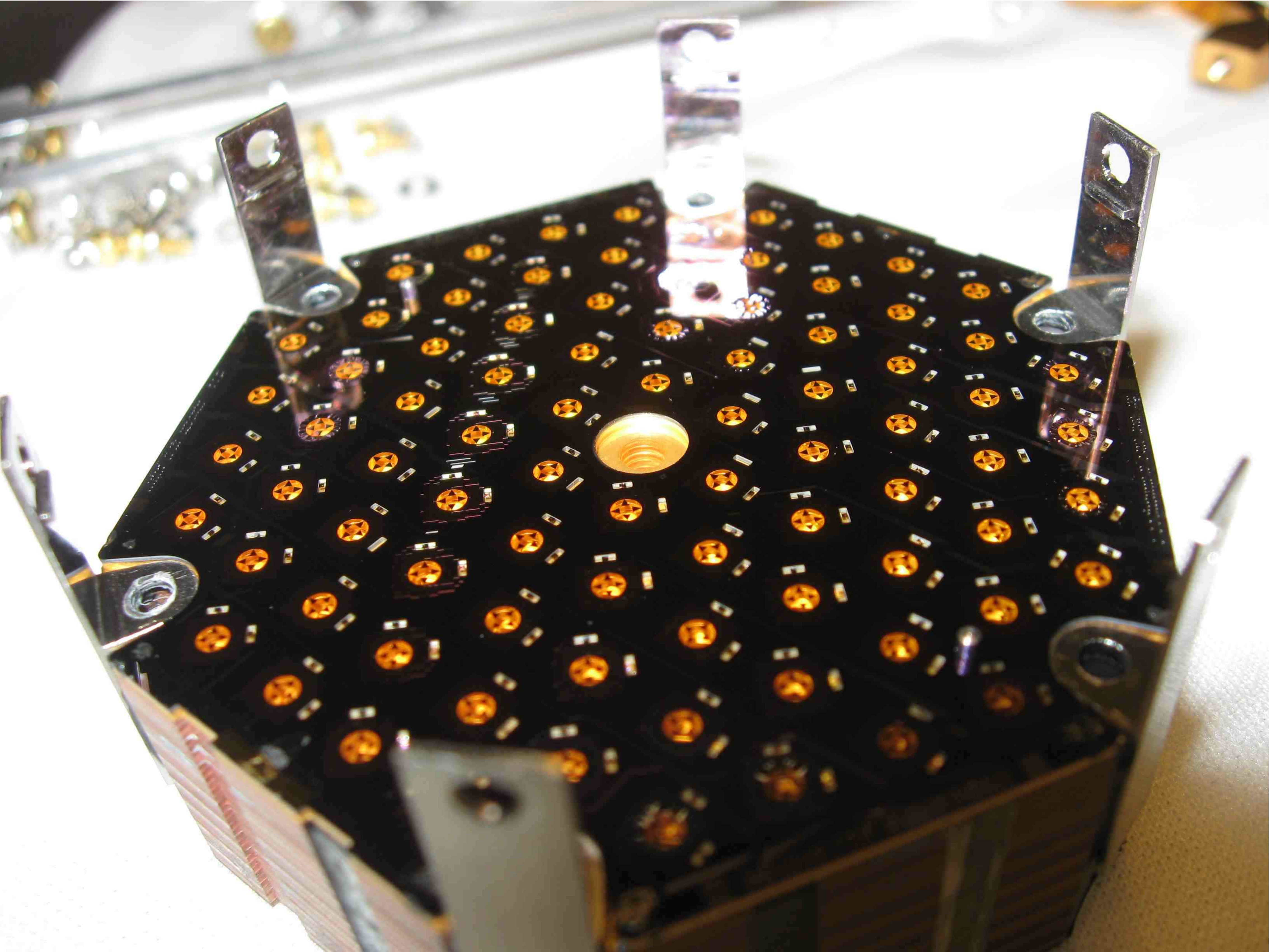}}
\caption{(Left) Schematic of one SPTpol 150 GHz pixel.  Power couples to an OMT, travels through coplanar waveguide to microstrip transition and microstrip, and is deposited on two released TES islands where it is dissipated by lossy gold meander.  Areas with white hashing are regions where the silicon substrate has been etched away leaving released silicon nitride.  (Right) One of the seven SPTpol 150 GHz detector arrays during module assembly, before installing the BS wafer.  The array is 2.3 inches across.}\label{pixel}
\end{center}
\end{figure}

Each module contains a detector array containing 84 dual-polarization pixels operated at $\sim$ 480 mK.  The arrays are monolithically fabricated by photolithography techniques on silicon wafers at NIST-Boulder.  The pixels are the result of development by the TRUCE collaboration\cite{Austermann_2009, Bleem_2009, Yoon_2009, Henning_2010, Hubmayr_2012a}.  Figure \ref{pixel} shows a schematic of a single pixel as well as one SPTpol detector array.  Power is coupled to a released orthomode transducer (OMT), which splits the light into two orthogonal polarization states.  The coupled power then travels down a coplanar waveguide (CPW) to microstrip transition, then through microstrip, and is eventually deposited on two transition edge sensor (TES) islands by a length of lossy gold meander.  The TES devices themselves are made of an aluminum manganese alloy.  TES devices operating in the middle of their superconducting transitions are extremely sensitive to small changes in incoming optical power.  We use a digital frequency-domain multiplexing readout system developed at McGill University\cite{Smecher_2012} to measure the change in detector temperature and therefore optical signal from the sky.  Superconducting microstrip carries the signal from the TES out from each pixel and to the array edges where four banks of 90 wire bonding pads exist.

\begin{figure}[t]
\begin{center}
\resizebox{0.5\textwidth}{!}{
\includegraphics[width=0.8\textwidth]{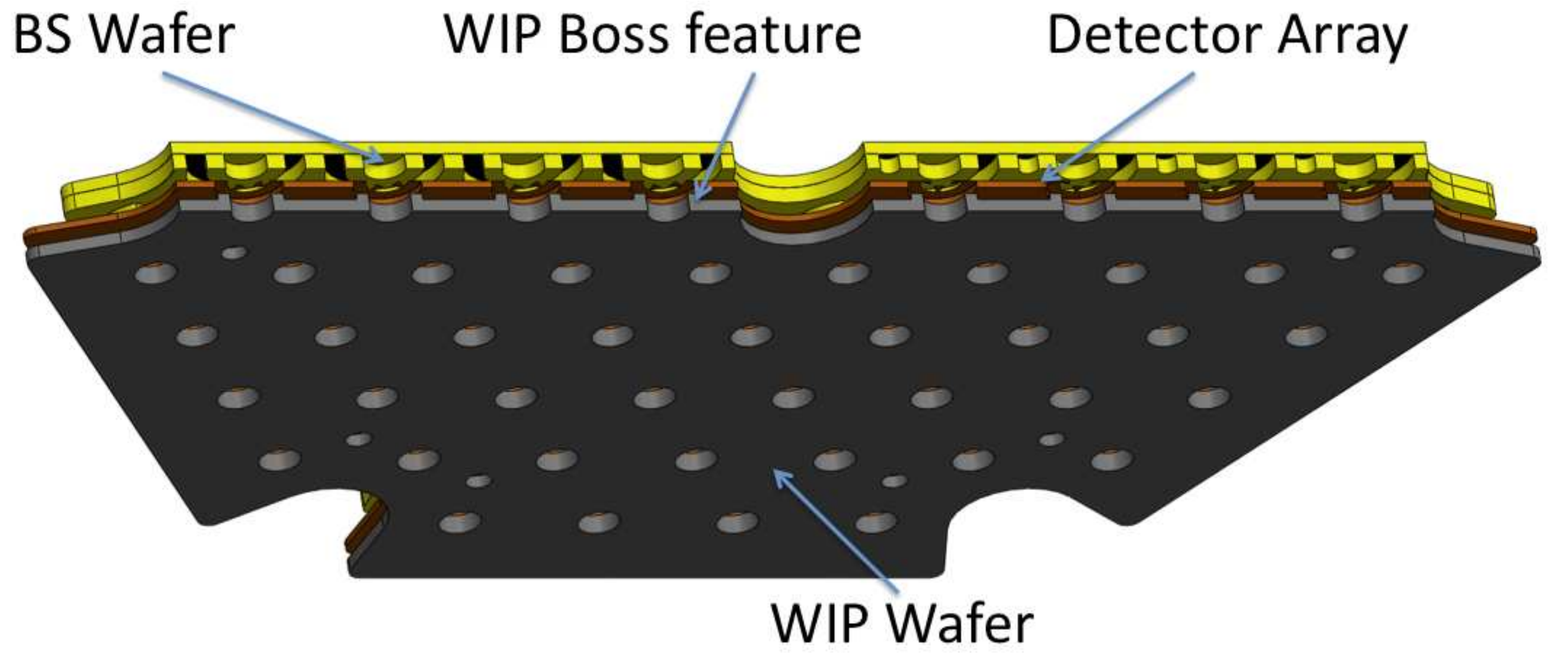}}
\caption{Expanded cross-sectional view of the detector wafer sandwich.  Boss features in the WIP slip into the back-etched cavities of each pixel OMT, extending the horn waveguide to the OMT and providing pixel-waveguide registration.  The BS wafer terminates the waveguides with $\lambda$/4 backshorts.}\label{sandwich}
\end{center}
\end{figure}

So that each detector array is sensitive to both Stokes Q and U polarization parameters, the orientation of an OMT is rotated by 45$^{\circ}$ with respect to its neighbors in an alternating pattern.  By installing the 150 GHz modules into the SPTpol focal plane rotated by 30$^{\circ}$ with respect to one another, we simultaneously measure Stokes Q and U with three sets of independent measurements.  These independent measurements allow for more tests to search for systematics in the data.

In addition to two optically-coupled TES devices per pixel, five pixels on each array also have a third TES, which is not connected to the OMT and is therefore a ``dark" device.  These dark TESs provide diagnostic checks for each array and can be used to test for non-OMT coupling power.  Including the dark devices, each array has a total of 173 detectors.  There are also two normal resistors on each array, which can be used to determine the array temperature.  

Each detector wafer is sandwiched between two ancillary silicon wafers that not only physically protect the detectors, but also help maximize in-band coupling, reduce out-of-band coupling, and minimize optical crosstalk.  Figure \ref{sandwich} shows an expanded cross-sectional view of the complete detector array sandwich.  The first ancillary wafer is a waveguide interface plate (WIP) located between the feedhorns and a detector array.  The WIP contains 250 $\mu$m tall boss features that continue the circular waveguide from the end of the feedhorn to within 25 $\mu$m of the lower surface of the released pixel OMTs.  The boss features neatly register within the back-etched cavities behind the OMTs and along with 1 mm wide slip fit alignment pins in the feedhorn array provide OMT-waveguide alignment to $\sim 25\,\mu$m.

\begin{figure}[h]
\begin{center}
\resizebox{0.75\textwidth}{!}{
\includegraphics[width=0.15\textwidth]{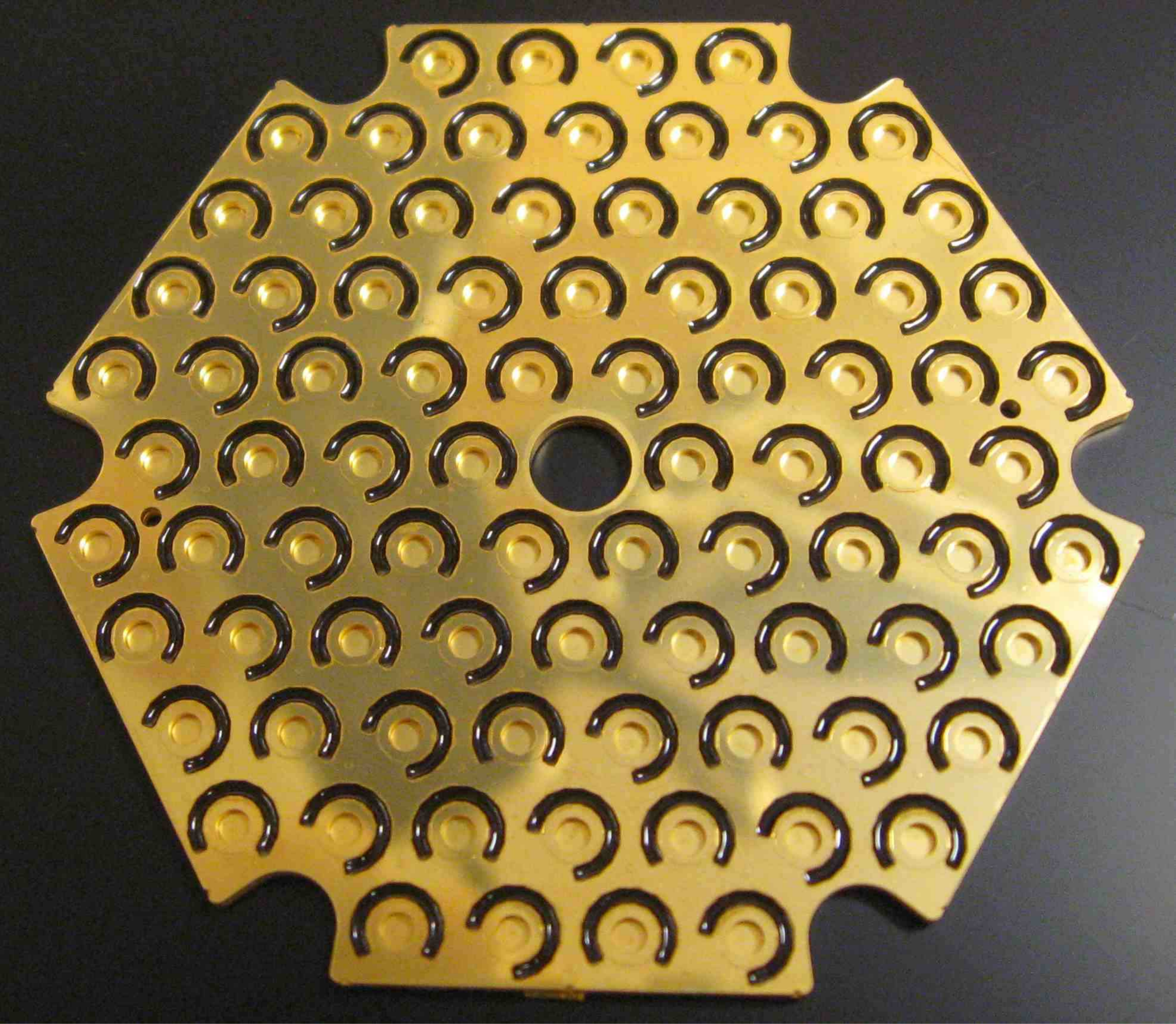}
\includegraphics[width=0.1475\textwidth]{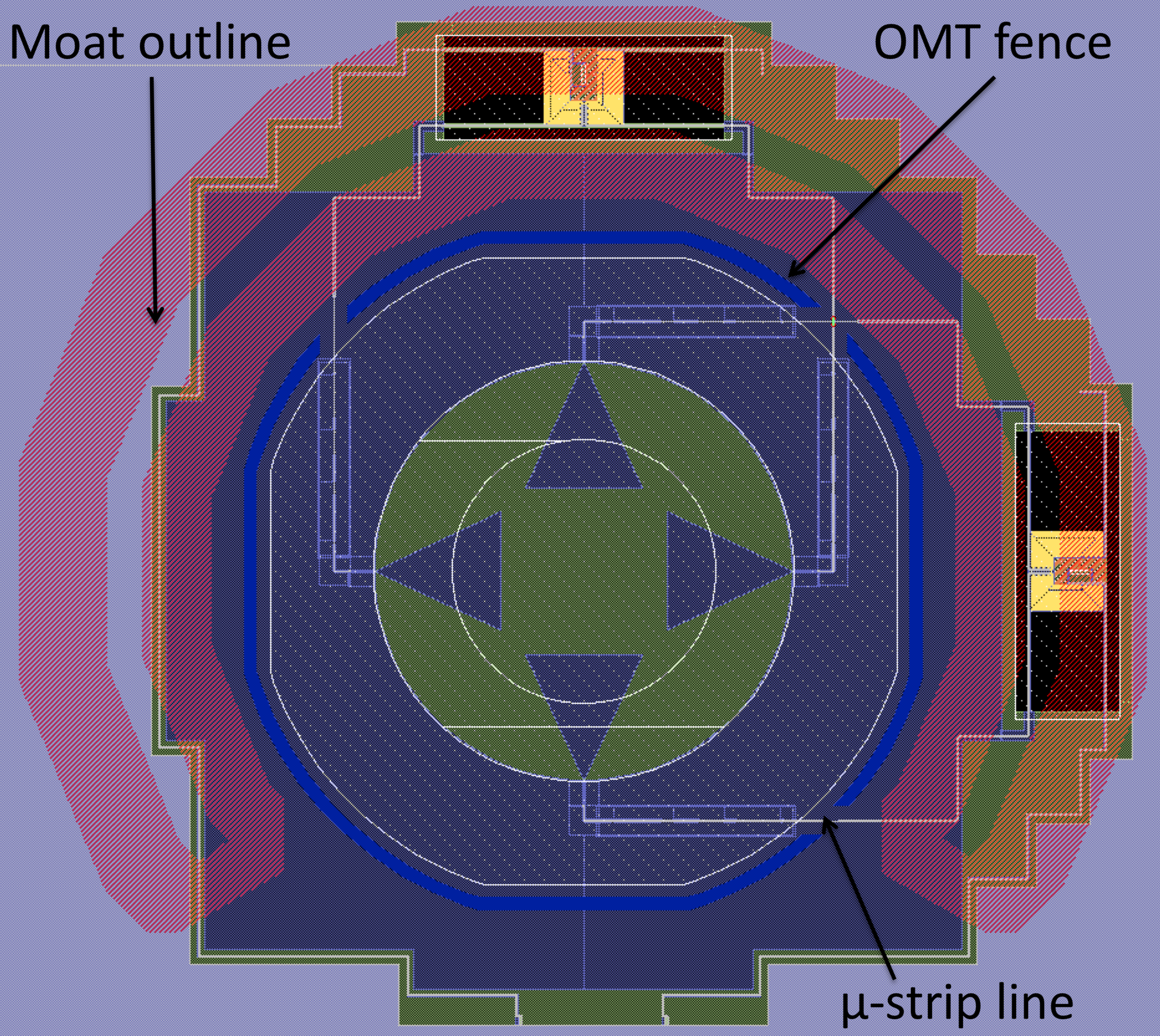}}
\caption{(Left) BS wafer with filled moats.  The moats alternately rotate by 45$^{\circ}$ to match the alternating rotations of the pixels.  (Right) A schematic of the BS moat and OMT fence layout with respect to pixel features.  The moats are aligned to be above each TES island, and the OMT fence closes the gap between the detector array and BS array except where microstrip lines are present.}\label{BS}
\end{center}
\end{figure}

The second ancillary silicon wafer is a $\lambda$/4 backshort (BS) wafer, shown in Figure \ref{BS}.  Not only does the backshort wafer terminate the waveguides, but it also contains ECCOSORB\cite{eccosorb} filled moats positioned above each TES island.  During detector development we found that prototype devices had a tendency to absorb high frequency out-of-band power coupling directly to the TES silicon nitride islands\cite{Henning_2010}.  This was exacerbated by using on-chip $\lambda$/4 stub filters to define the bandpass so that no filtering was applied before light coupled through the feedhorns.  By placing black material above or below the TES islands, however, the amount of out-of-band power picked up by the devices reduced by factors of 2-3.  SPTpol pixels use low-pass filters in front of the feedhorns to define the band, which removes high frequency out-of-band light.  We nevertheless installed the ECCOSORB moats to absorb any stray non-OMT coupling light that may regardless find its way to our detectors.  

Additionally, the BS wafer rests on top of the detector wafer in the sandwich.  To not crush the detectors, it is stood off from the detector array by 25 $\mu$m tall silicon dots scattered across the BS wafer.  This creates a 25 $\mu$m gap between the pixel OMTs and the backshort cavities, which allows light to leak out of the waveguide and couple to the TES's directly instead of through the OMT and microstrip.  To reduce this potential source of optical crosstalk, the backshort cavities are surrounded by 25 $\mu$m tall silicon fences to close as much of the gaps as possible.  The only remaining gaps allow the microstrip underneath to leave the cavity and continue to the devices themselves.  Figure \ref{BS} contains an illustration of the moat geometry and OMT fencing with respect to the pixel layout.

\subsection{Mounting Hardware}\label{Hardware}
\begin{figure}[h]
\begin{center}
\resizebox{0.6\textwidth}{!}{
\includegraphics[width=1.5\textwidth]{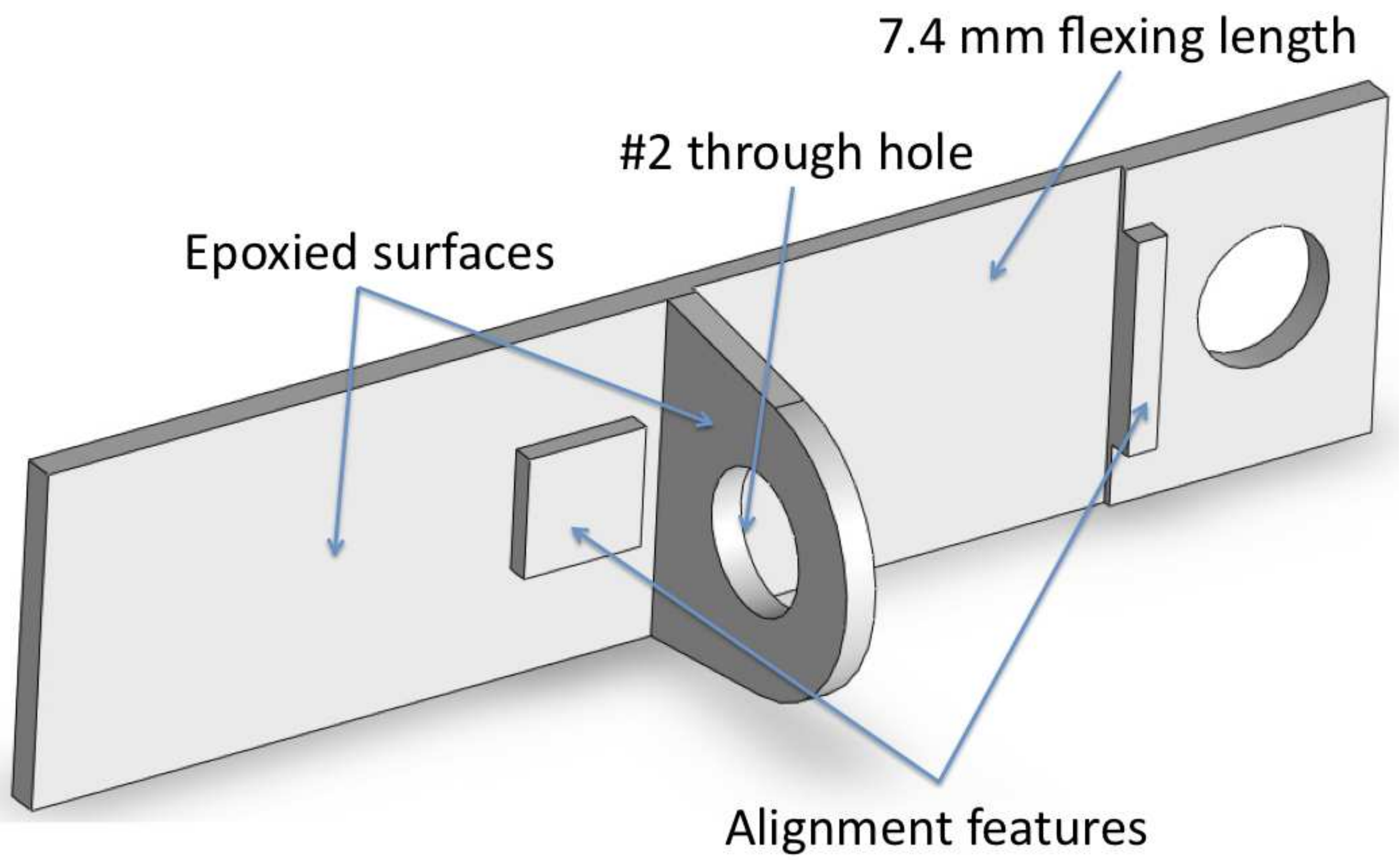}
\includegraphics[width=1.0\textwidth]{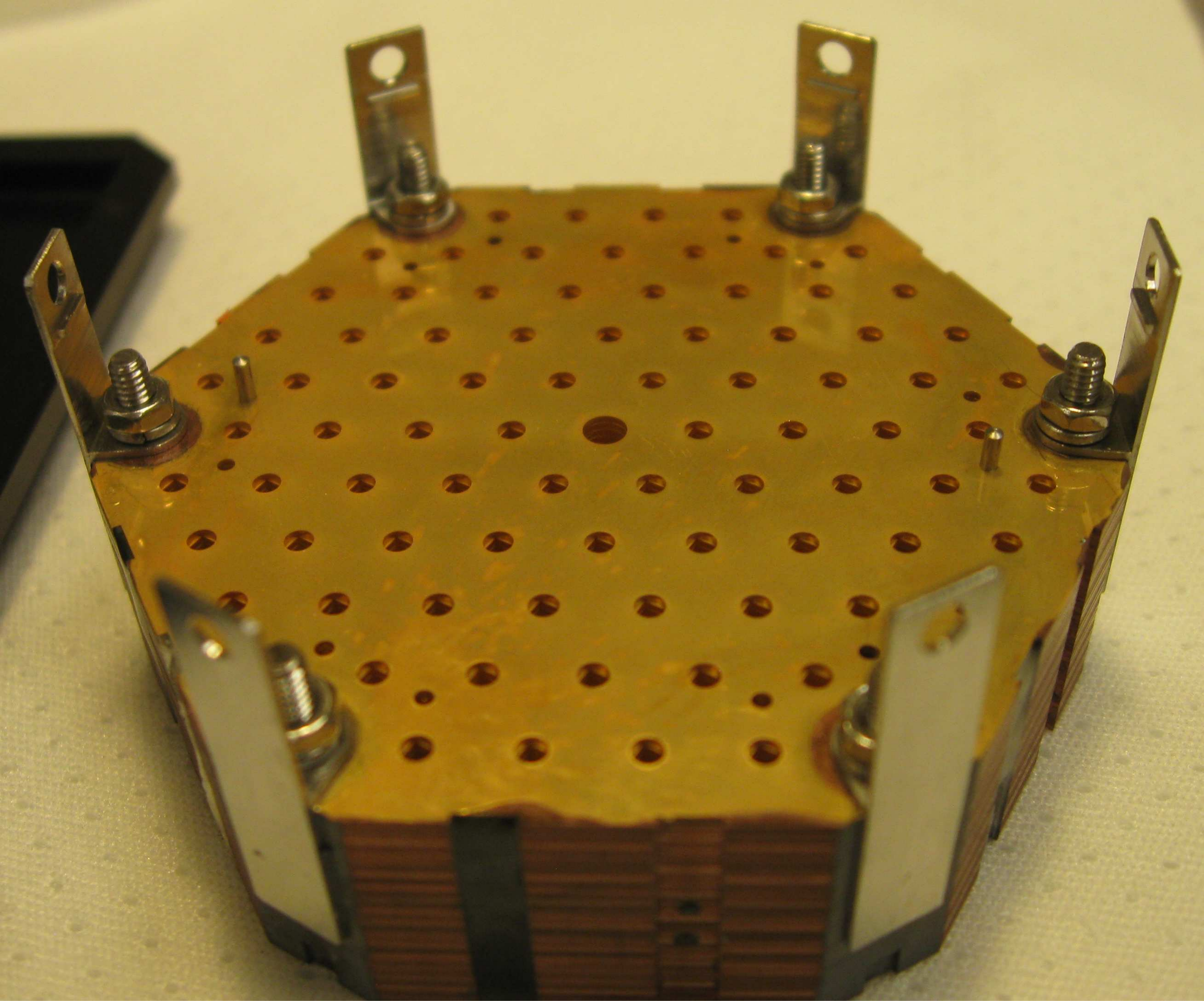}}
\caption{(Left) Flexible invar tabs used to connect the silicon feedhorn arrays to the metal mounting structures in the 150 GHz modules.  Each tab is 0.5 mm thick, while the flexing portion is 0.4 mm thick to reduce stress near the epoxy joint.  The flexible portion of the tab, between the registration features, is 7.4 mm long.  (Right) The tabs installed on a feedhorn array prior to module assembly.}\label{tabs}
\end{center}
\end{figure}

While the feedhorn and detector sandwich arrays are both made of silicon, the rest of the focal plane and modules are comprised of copper and aluminum.  Since these metals contract $\sim$ 20 times more than silicon when cooling from 300 to 4 K, we need a mounting scenario that prevents fracturing the detector arrays, the feedhorns, or both.  We use six flexible T-shaped tabs made of invar mounted to the six corners of the feedhorn array, shown in Figure \ref{tabs}.  As the copper mounting components in the module shrink with respect to the silicon upon cooling, the tabs flex to absorb the size difference.  Invar has a linear coefficient of thermal expansion that is only twice that of silicon, a negligible difference over the size scales in question resulting in differential contraction between the invar and silicon of $< 4\,\mu$m.

The tabs are permanently adhered to a silicon feedhorn array using a thin layer of stycast 2850\cite{stycast} on the corner walls of the array.  Alignment features in both the feedhorn array and the tabs ensure the tabs are installed in the proper position.  Prototype tests using copper tabs revealed that differential contraction between copper and silicon produced enough stress in the stycast epoxy to weaken the glue joint, which caused the tabs to fall off upon multiple thermal cycles.  We optimized the shape of the invar tabs to reduce stress on the epoxy joint, making the flexible portion 7.4 mm long.  While the tab is 0.5 mm thick, the flexible portion of each tab is thinned to 0.4 mm to direct the point of highest stress in the tab away from the epoxy joint.  Additionally, the mounting hardware to which the tabs attach is oversized at 300 K.  Upon cooling to 4 K, the distance between two opposite tab mounting points equals the width of the feedhorn array, so that at operating temperatures where the epoxy is most brittle there is zero flexing in the tabs and negligible stress to the epoxy joints.  

While this invar/epoxy configuration proved robust to multiple thermal cycles, the invar-silicon joint must remain intact to ensure pointing and beam properties are consistent throughout an observing season.  For additional security, \#2-56 threaded invar posts were epoxied into through holes in each of the six corners of the feedhorn arrays, again using stycast as the epoxy.  A mounting tab protruding from the invar tabs slips over these threaded mounting posts, is epoxied into place, and a nut screwed onto the post provides vertical pressure.  With the added nut-and-bolt mounting, we ensure that even in the unlikely event that one of the lateral epoxy joints fail, the tab is still securely in place.  The tabs can be seen installed on a feedhorn array in Figure \ref{tabs}.

\begin{figure}[h]
\begin{center}
\resizebox{1.0\textwidth}{!}{
\includegraphics[width=2.0\textwidth]{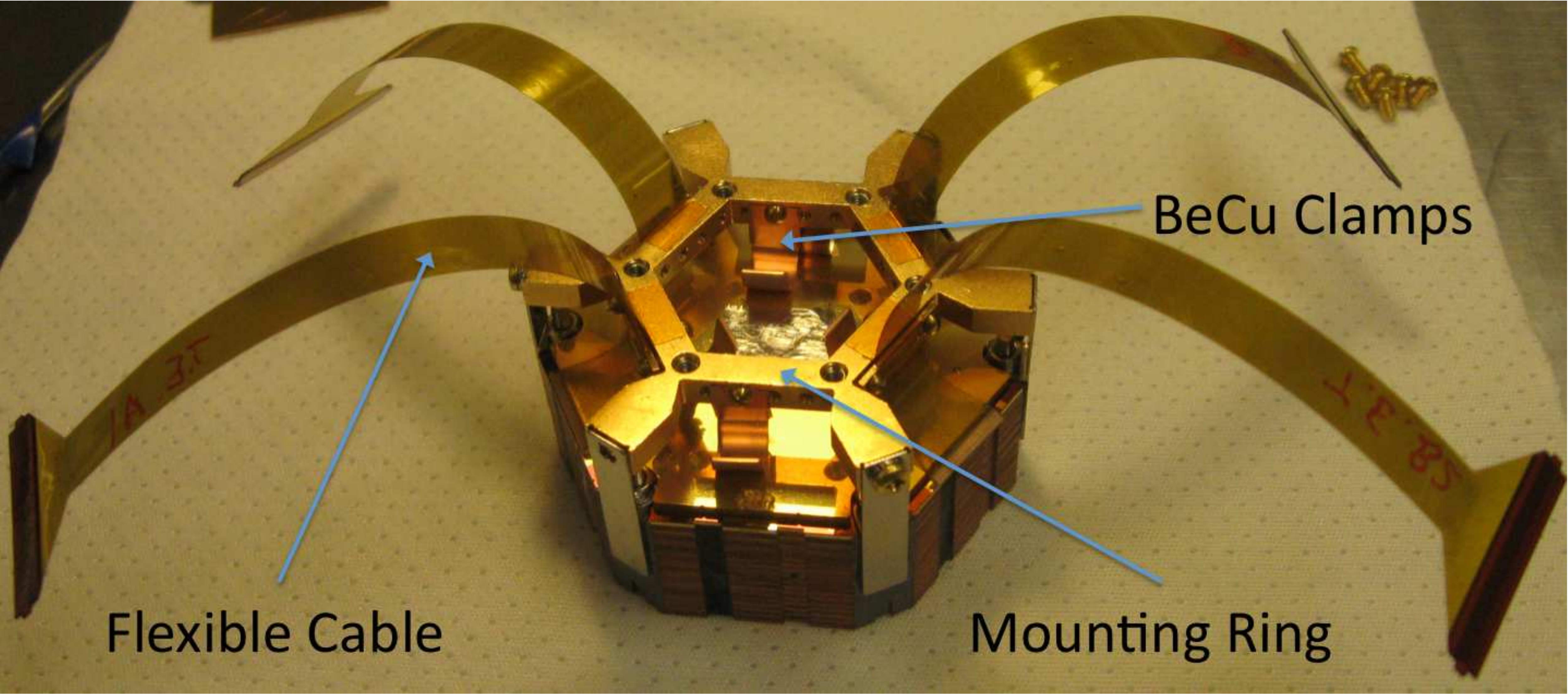}
\includegraphics[width=1.0\textwidth]{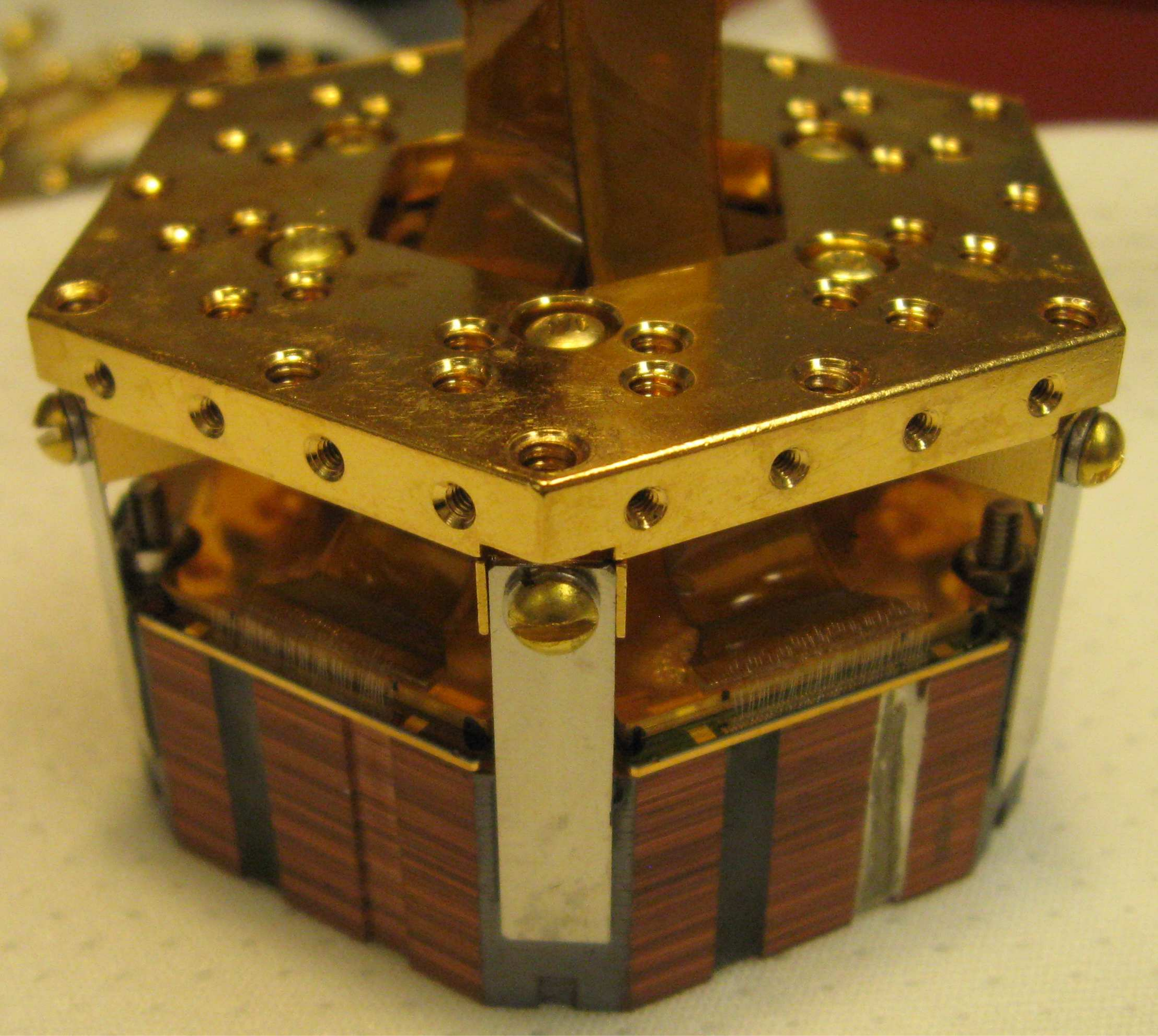}}
\caption{(Left) Mounting ring with berylium copper clamps installed in a module.  The invar tabs screw into the six feet in the ring, while four flexible cables are clamped down to the outer walls of the ring.  The berylium copper clamps supply vertical pressure to the detector wafer stack.  (Right) The interface plate installed on top of the mounting ring with its many mounting holes.  The flexible readout cables snake underneath the plate and come out the center of the module.}\label{mounting}
\end{center}
\end{figure}

The invar tabs attach to a star-shaped copper mounting ring, shown in Figure \ref{mounting}.  The ring serves multiple functions in the module.  First, it acts as the point to which the invar tabs, and therefore the feedhorn array, can mount.  The mounting ring shape is designed to provide tool access, both for assembling various components of the module as well as for wire bonding the detector arrays to flexible readout cables (see Section \ref{Readout}).  Each of the six sides of the mounting ring has \#0-80 tapped mounting holes, which are used to screw down strain relief bars for the readout cables as well as to hold beryllium copper clips that vertically constrain the detector wafer sandwich.  Second, the mounting ring functions as a heat strap between the silicon feedhorns and detectors and the focal plane millikelvin plate.  The thermal conductivity of invar is three orders of magnitude lower than copper below 1 K, so the invar tabs play an insignificant role in cooling the silicon.  Instead, cooling is achieved through the beryllium copper vertical clamps, as well as the RF skirts (see Section \ref{Shielding}) that mount to a copper interface plate which itself attaches to the mounting ring.  This interface plate, also shown in Figure \ref{mounting}, provides a location for more structure mounting.  In addition to the RF shields, the readout board sub-assembly attaches here.  The interface plate is also the surface that bolts directly to the focal plane millikelvin plate.

\subsection{RF Shielding}\label{Shielding}
\begin{figure}[h]
\begin{center}
\resizebox{0.55\textwidth}{!}{
\includegraphics[width=1.25\textwidth]{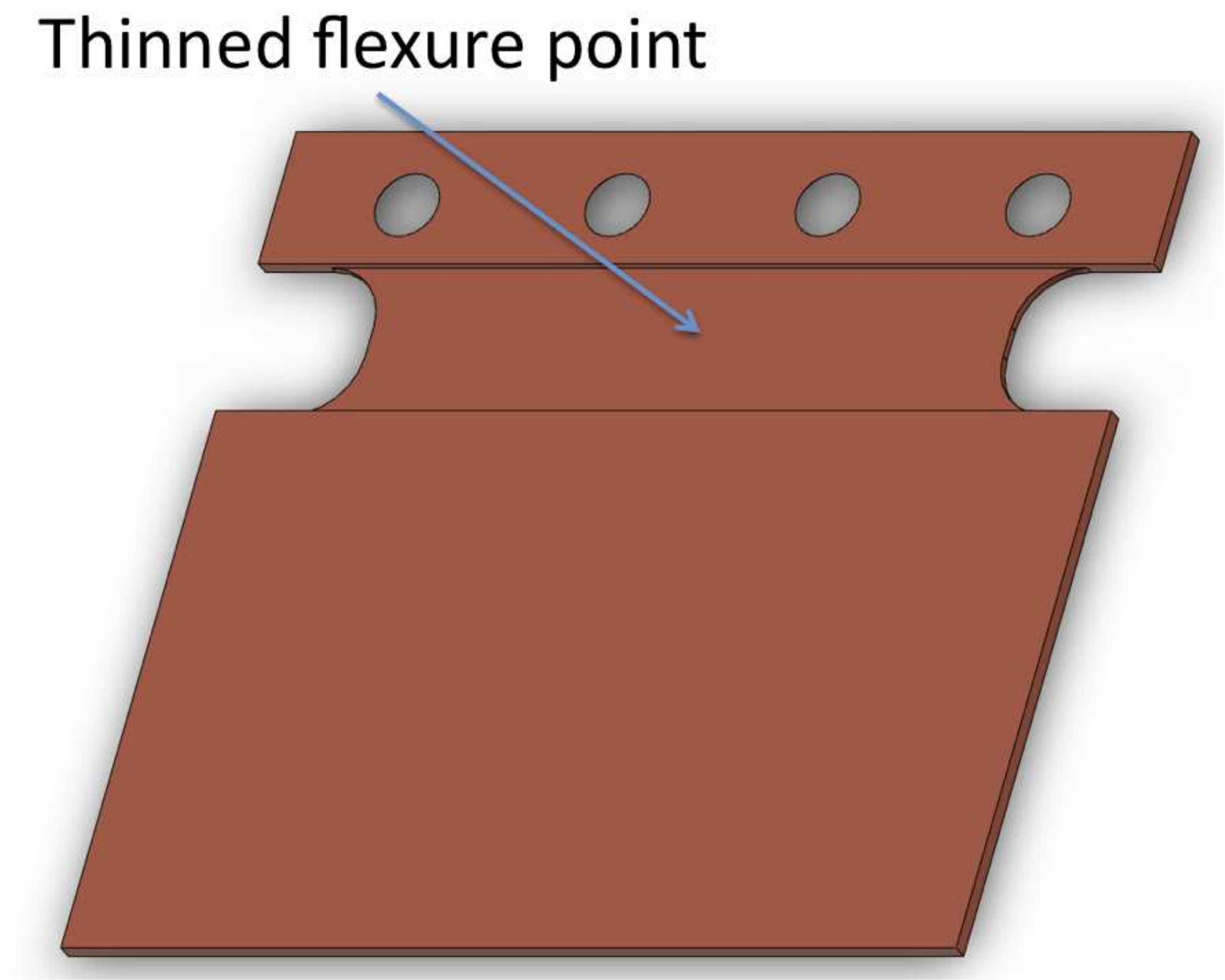}
\includegraphics[width=1.0\textwidth]{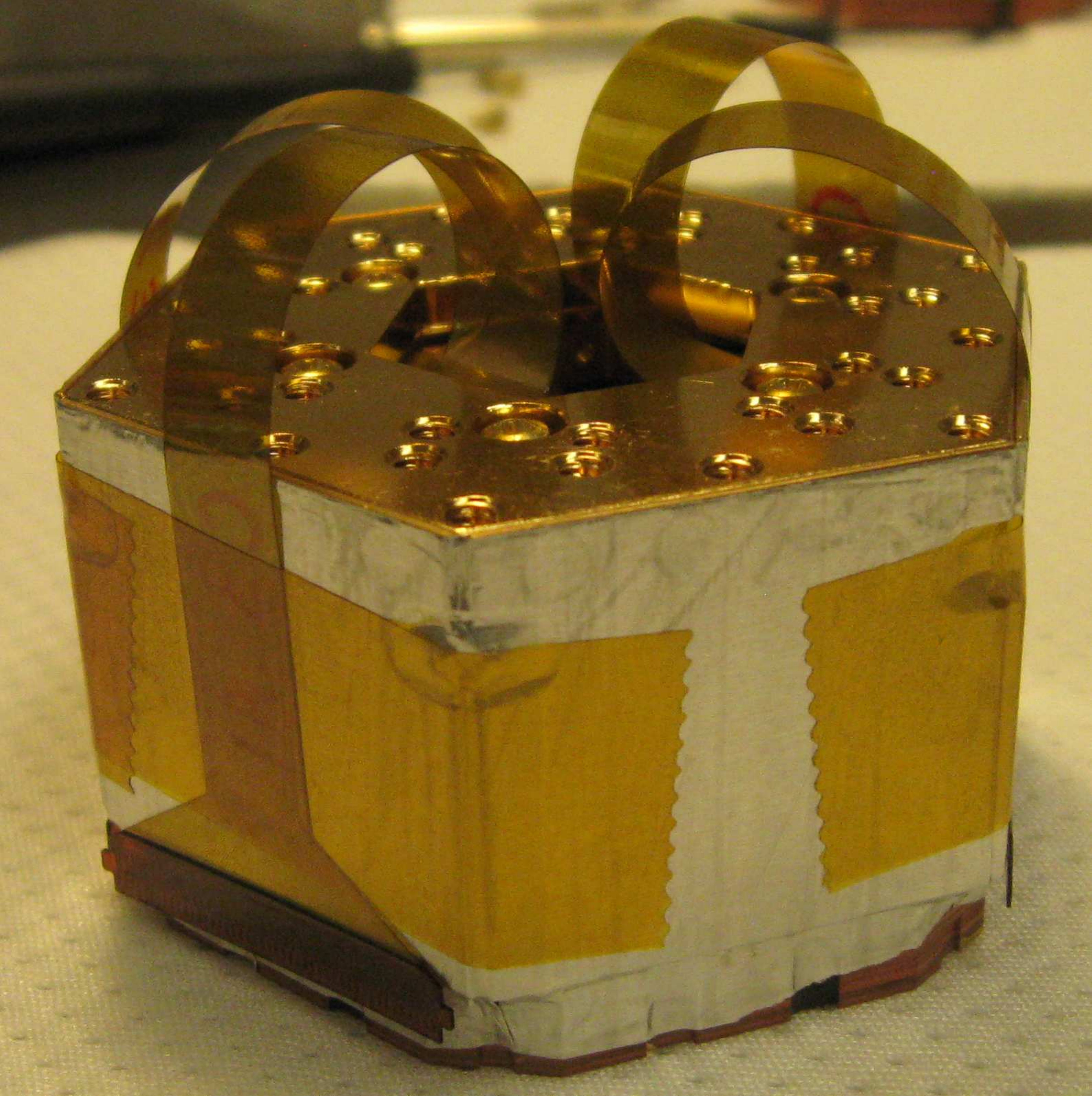}}
\caption{(Left) Image of RF skirts, showing flexure portion.  (Right) Skirts installed with final RF tape applied as the module is prepared for shipment to the South Pole.}\label{skirts}
\end{center}
\end{figure}

RF light can couple directly to our readout electronics, acting as a source of extra noise in our system.  As an independent camera, each module must be RF-tight when installed in the focal plane, only allowing light to enter the module through the feedhorns, before which capacitive metal mesh low-pass filters define the upper edge of our observing band and the high-pass filter of the feedhorn waveguide defines the lower edge of the band.  Having long flexible tabs between the silicon feedhorns and metal mounting hardware leaves the sides of the module completely open.  To close these large gaps, thin copper skirts shown in Figure \ref{skirts} attach to the sides of the interface plate and come down the length of the module, to roughly a half inch from the top of the silicon feedhorn array.  

These RF skirts also act as a mechanical shield for the delicate wire bonds on the detector wafers, only a few millimeters away from the edge of the feedhorn array.  Aluminum tape seals the small gaps left between the skirts, adhering to the sides of the feedhorn array.  Figure \ref{skirts} shows a complete module assembly, with RF skirts installed and aluminum tape applied.  Since the feedhorns are metal-plated, there is a contiguous conducting surface all the way from the feedhorn apertures to the focal plane millikelvin plate when the module is installed ensuring light can only enter the system through the feedhorns.

As mentioned previously, the RF skirts also cool the silicon feedhorn and detector arrays.  The interface plate to which the skirts attach is the same width as the silicon arrays at 300 K.  As the module cools, the interface plate shrinks with respect to the silicon array.  The differential contraction squeezes the RF skirts tighter against the walls of the feedhorn array, which provides pressure points through which to cool the silicon.  Too much pressure on the silicon from the skirts could chip or crack it, however.  To avoid this outcome, the RF skirts have a horizontal band across them that has been thinned to act as a flexure point, reducing the total pressure on the feedhorn array while allowing the skirts to supply enough pressure to effectively cool the silicon.

\subsection{Passive Readout Electronics}\label{Readout}
To reduce loading on the millikelvin stage, the detectors are read out with a digital frequency-domain multiplexing (DfMux) readout system\cite{Smecher_2012}.  In a frequency-domain system, each detector is placed in series with an inductor $L$ and capacitor $C$ making an $RLC$ resonance circuit.  We AC bias many detectors simultaneously with a ``comb" of bias frequencies sent along a single pair of wires.  Given the $L$ and $C$ in series with a detector, each detector only sees the bias tone to which its resonance circuit is tuned.  The signals are then sent to a series array of 100 superconducting quantum interference devices (SQUIDs) followed by a low-noise amplifier cooled to 4 K, which together amplify the detector signals.  The SPTpol DfMux system uses a multiplexing factor of 12x, so 12 detectors (6 pixels) are biased with one set of wires and one SQUID series array.

While the SQUID arrays are cooled to 4 K, the passive $LC$ resonance circuits reside with the detector arrays and are therefore cooled to $\sim 480$ mK.  Printed circuit boards containing inductors and capacitors that define the resonance frequency of each detector (LC boards) are connected via aluminum supports to the 150 GHz modules.  Each board has 12 photolithographed chips of eight 22 $\mu$H inductors fabricated at NIST-Boulder, as well as capacitors stacked to achieve the requisite capacitance.  Therefore, each board can define the resonance frequencies for up to 96 different detectors.  In practice, only 90 channels on each of two LC boards are needed to define the resonance channels for all the detectors in a single 150 GHz module.  

\begin{figure}[t]
\begin{center}
\resizebox{0.65\textwidth}{!}{
\includegraphics[width=1.0\textwidth]{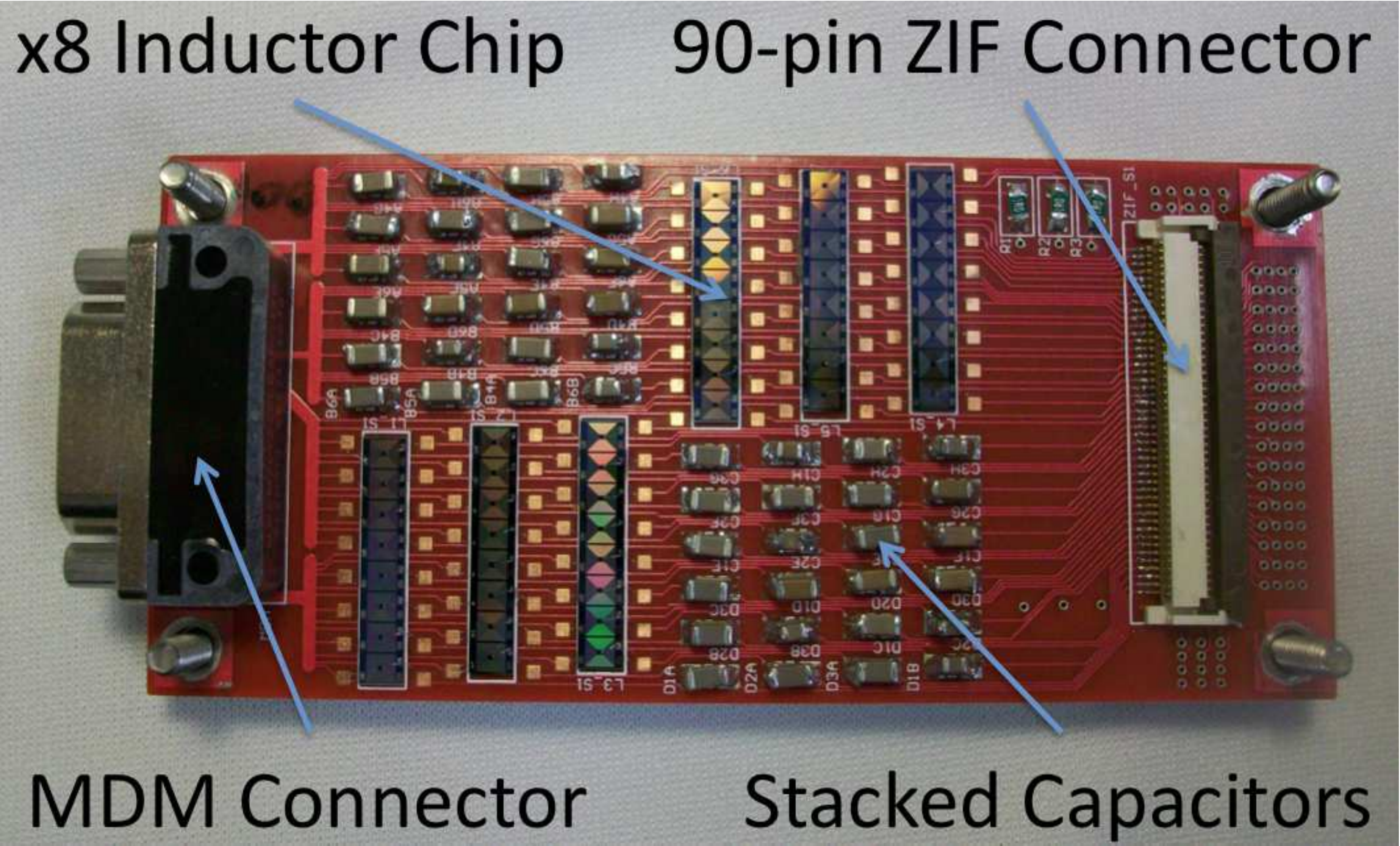}}
\resizebox{0.65\textwidth}{!}{
\includegraphics[width=1.0\textwidth]{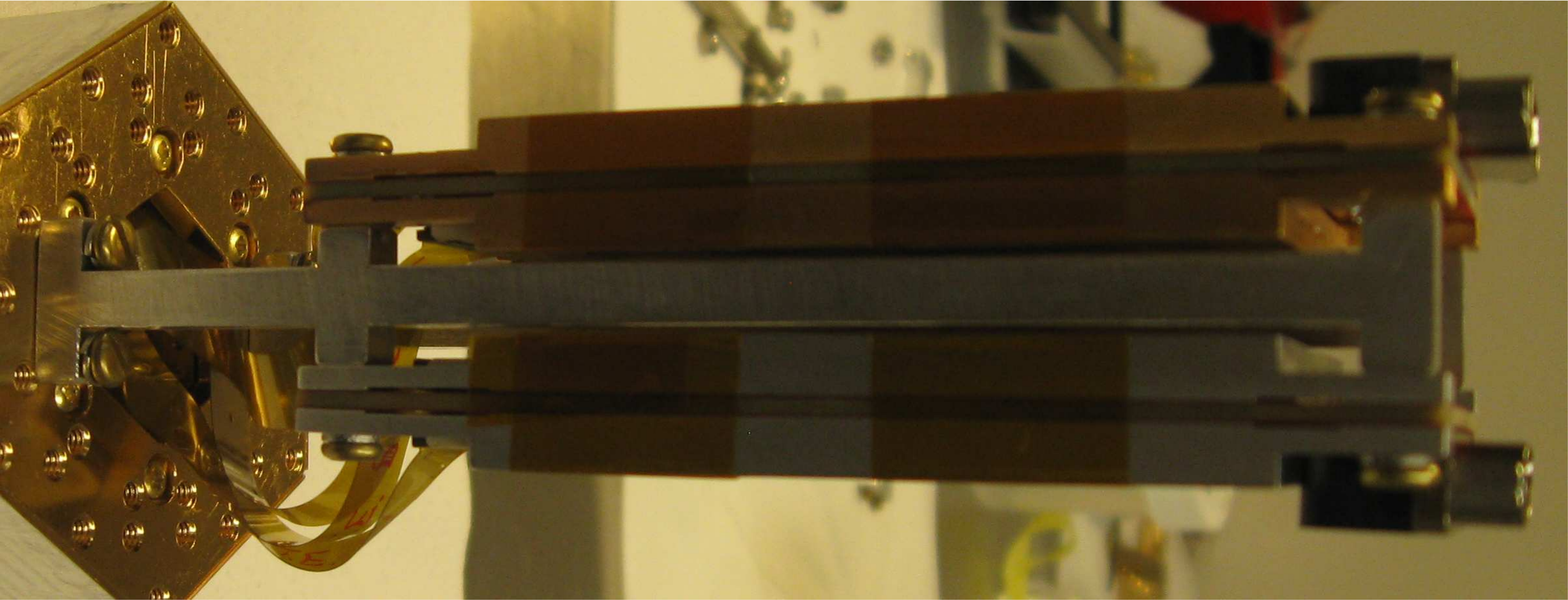}}
\caption{(Top) Populated LC board.  (Bottom) Two LC boards with protective shielding in place and mounted to a module.}\label{LC}
\end{center}
\end{figure}

To protect the delicate inductor chips and wire bonds connecting them to the circuit boards, thin aluminum shields are placed over the LC boards.  Since the fiberglass body of the circuit board is black at infrared wavelengths, the aluminum shields also reduce the surface area of the boards that could otherwise terminate warm stray light in the experiment cryostat, reducing the parasitic heat load on the millikelvin refrigerator.

The LC boards are physically displaced from the detector arrays by several inches in each module.  To bridge the gap, we use the flexible circuit cables mentioned previously.  The cables contain 90 copper traces printed on a polyimide substrate, which are 47 $\mu$m wide and have 94 $\mu$m center-center pitch.  The traces are tinned to be superconducting to reduce parasitic resistance in the $RLC$ circuits.  A final polyimide overlay layer protects the traces and reduces the chance of electrical shorts between the cables and module hardware.  The cables are rubber cemented into position and strain-relieved with clamping bars.  One end of the cable has a series of bare copper bond pads.  We wire bond from the detector array bond pads to the flexible cable pads using 1 mil thick aluminum bonding wire.  The other end of the flexible cable has 90 zero insertion force (ZIF) contacts.  This end of the cable is then easily attached to an LC board by plugging into a ZIF connector on the board.  Two cables plug into each LC board, which are populated on both the front and back sides.
 
\section{Dark Properties}\label{Dark}
Several detector properties can be obtained by measuring the current through a device as a function of applied voltage bias (an $IV$ curve)\cite{Austermann_2009,Henning_2010}.  Please refer to the cited references for examples of this measurement technique.  Lab observations of device IV curves provided measurements of dark properties for five of the seven deployed detector arrays.  We also fitted the width and height of detector $RLC$ resonance peaks above and below device superconducting transitions to obtain normal resistances $R_N$ and parasitic resistances in series with detectors, respectively.  There are systematics at the level of 10-20\% for these resonance fits due to calibration uncertainties in the lab, however devices observing the sky can be well calibrated to a celestial source.  Table \ref{dark_properties} contains a summary of the results for each wafer.  Means and standard deviations are provided, as well as the number of detectors measured.  Normal resistances and measured saturation powers are corrected for in-series parasitic resistance, and only devices with both $RLC$ resonance fits and $IV$ curve measurements are included.  Four of the wafers show good uniformity in superconducting critical temperature $T_c$, normal resistance, and saturation power $P_{sat}$.  Wafer D4, however, does have a clear bimodal distribution of devices, split into ``Low" and ``High" columns based on whether devices have saturation powers above or below 25 pW.  Disregarding the ``High" outlier distribution in wafer D4, the average dark properties across the five lab-tested arrays are $T_c = 468.9 \pm 12.4$ mK, $R_N = 1.2 \pm 0.2\,\Omega$, and $P_{sat} = 20.6 \pm 2.9$ pW.  Including all measured devices on the five arrays the averages are $T_c = 478.0 \pm 28.6$ mK , $R_N = 1.2\pm 0.2\,\Omega$, and $P_{sat} = 22.5\pm 5.7$ pW.

\begin{table}
\begin{center}
\begin{tabular}{c||c|c||c|c||c|c}
Wafer & \multicolumn{2}{c||}{$T_c$ (mK)} & \multicolumn{2}{c||}{$R_N$ ($\Omega$)}  & \multicolumn{2}{c}{$P_{sat}$ (pW)} \\
\hline
C1 & $476.8 \pm 2.0$ & 87 & $1.1 \pm 0.1$ & 76 & $20.9 \pm 3.8$ & 74\\
C3 & $463.3 \pm 7.1$ & 132 & $1.2 \pm 0.2$ & 134 & $19.6 \pm 2.4$ & 134 \\
C4 & $467.1 \pm 3.2$ & 114 & $1.2 \pm 0.2$ & 123 & $21.4 \pm 2.5$ & 122 \\
C5 & $467.6 \pm 2.2$ & 131 & $1.2 \pm 0.2$ & 131 & $21.0 \pm 3.0$ & 131 \\
D4 Low & $478.2 \pm 34.0$ & 48 & $1.3 \pm 0.3$ & 48& $19.9 \pm 2.4$ & 48 \\
D4 High & $538.9 \pm 32.3$ & 76 & $1.0 \pm 0.1$ & 76 & $34.9 \pm 4.3$ & 76 \\
Average (No D4 High) & $T_c = 468.9 \pm 12.4$ & 512 & $R_N = 1.2 \pm 0.2$ & 512 & $P_{sat} = 20.6\pm 2.9$ & 509 \\
Average (With D4 High) & $T_c = 478.0 \pm 28.6$ & 588 & $R_N = 1.2 \pm 0.2$ & 588 & $P_{sat} = 22.5\pm 5.7$ & 585 \\
\end{tabular}
\vspace{0.05in}
\caption{Summary of in-lab detector dark tests for five of the seven deployed detector wafers.  Wafer D4 has a clear bimodal distribution in device properties and has been split into ``D4 Low" and ``D4 High."  Means and standard deviations are given as well as the number of devices tested for each wafer and property.}\label{dark_properties}
\end{center}
\end{table}

During detector development, early device designs exhibited high values for $\alpha = d\log R$/$d\log T$, a dimensionless ratio that quantifies the sharpness of a device's superconducting transition.  Higher values for $\alpha$ means a device is more sensitive to changes in temperature, (the loop gain of the device is proportional to $\alpha$ in the device transition), but also means the device can become unstable when biased too deep in the transition.  Indeed, our early devices were going unstable at operating bias points of $\sim 0.8\,R_N$, near typical operating points for our 150 GHz detectors\cite{George_2012}.  We tested several TES geometries to find an appropriate transition shape and deposited a non-superconducting metal, palladium gold (``bling"), around each TES to increase the detectors' time constants.  We found that having a solid bar for the TES and extending the bling over the edge of the microstrip leads heading to the TES and into the TES region itself lowered $\alpha$ sufficiently while keeping 150~GHz detector loop gains in transition at an acceptable level.  A companion paper in these proceedings\cite{Sayre_2012} discusses SPTpol 90 GHz pixel development and provides further detail about how the TES geometry affects $\alpha$.

\section{Optical Properties}\label{Optical}
\begin{figure}[t]
\begin{center}
\resizebox{1.0\textwidth}{!}{
\includegraphics[width=1.0\textwidth]{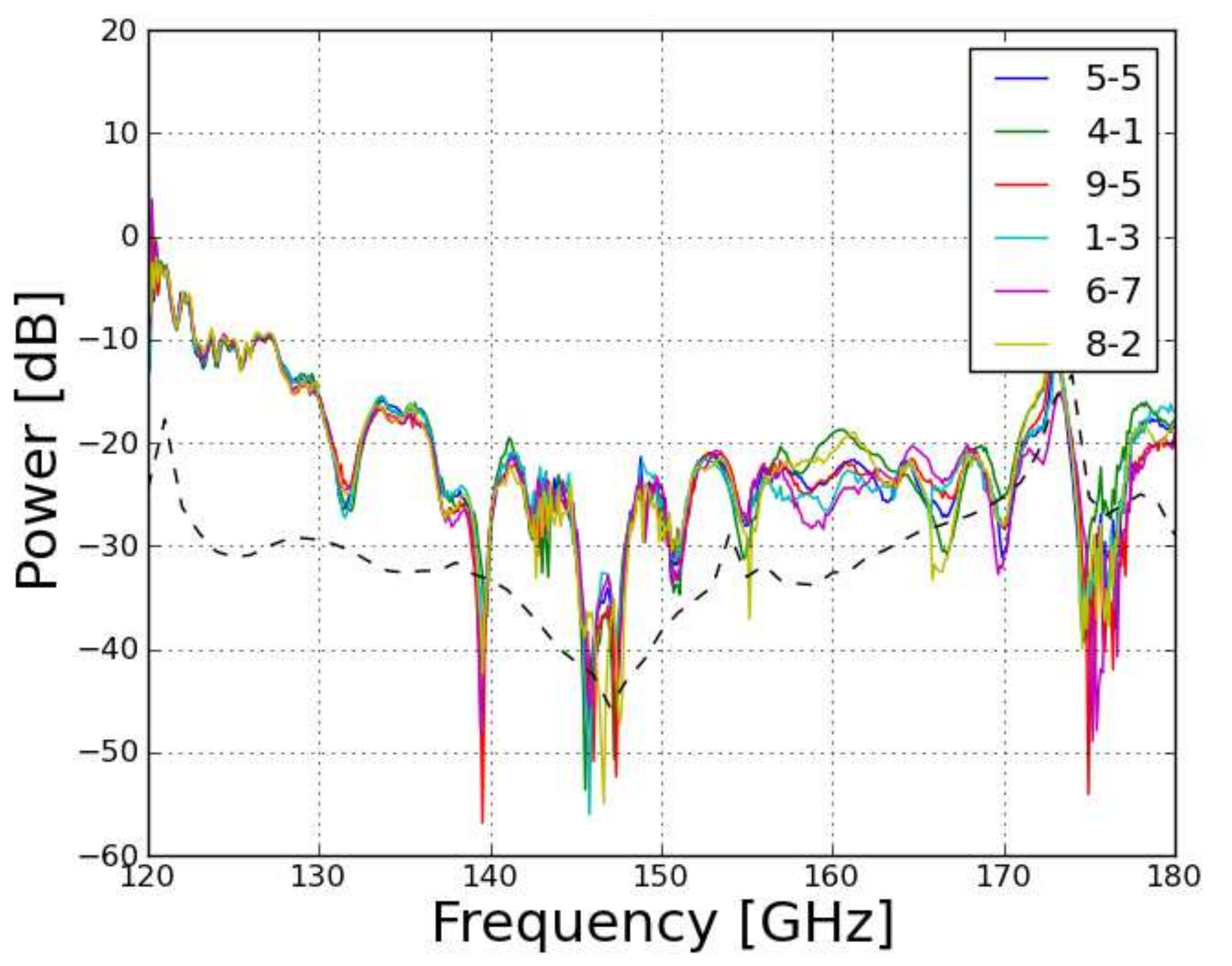}
\includegraphics[width=1.0\textwidth]{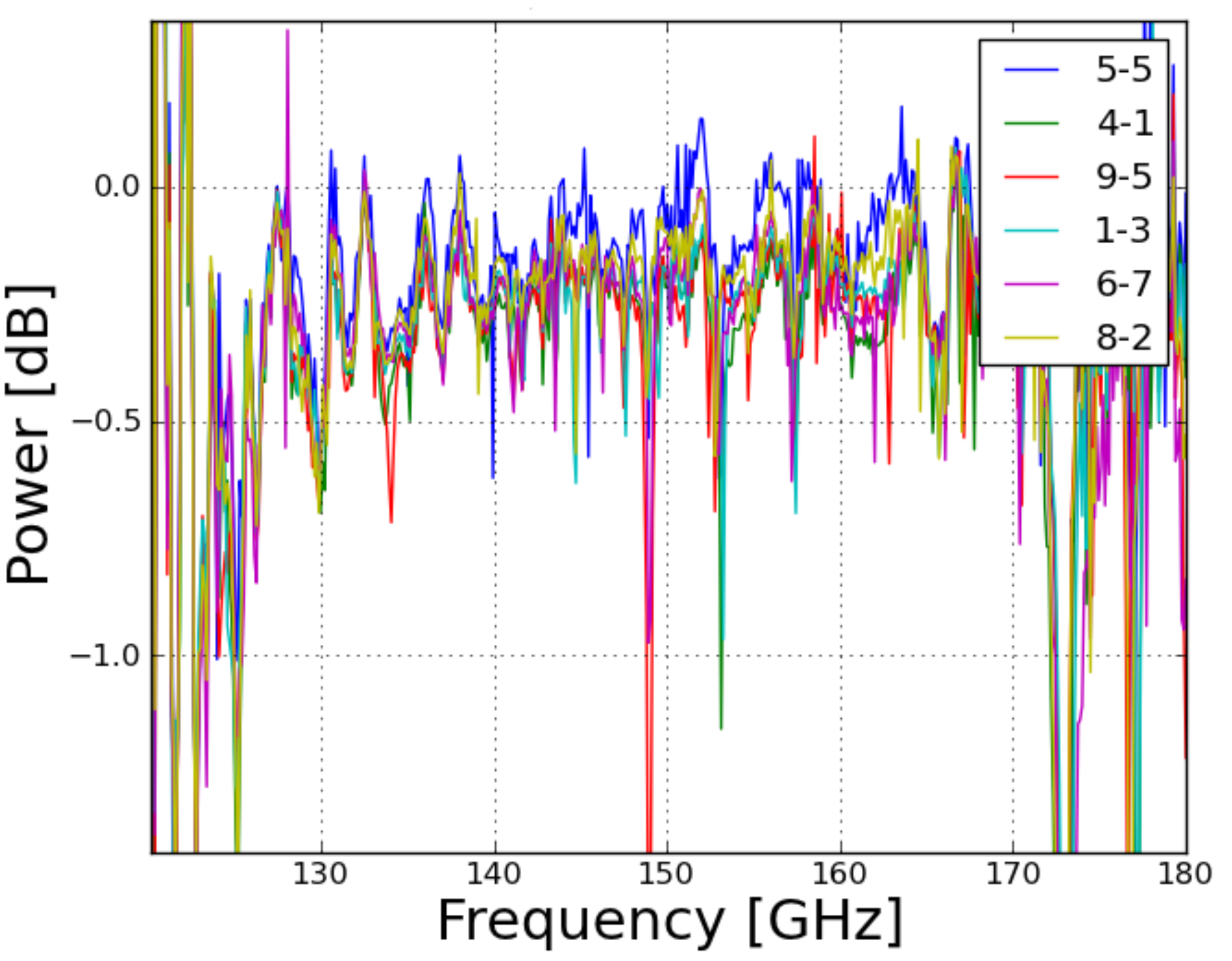}
\includegraphics[width=1.025\textwidth]{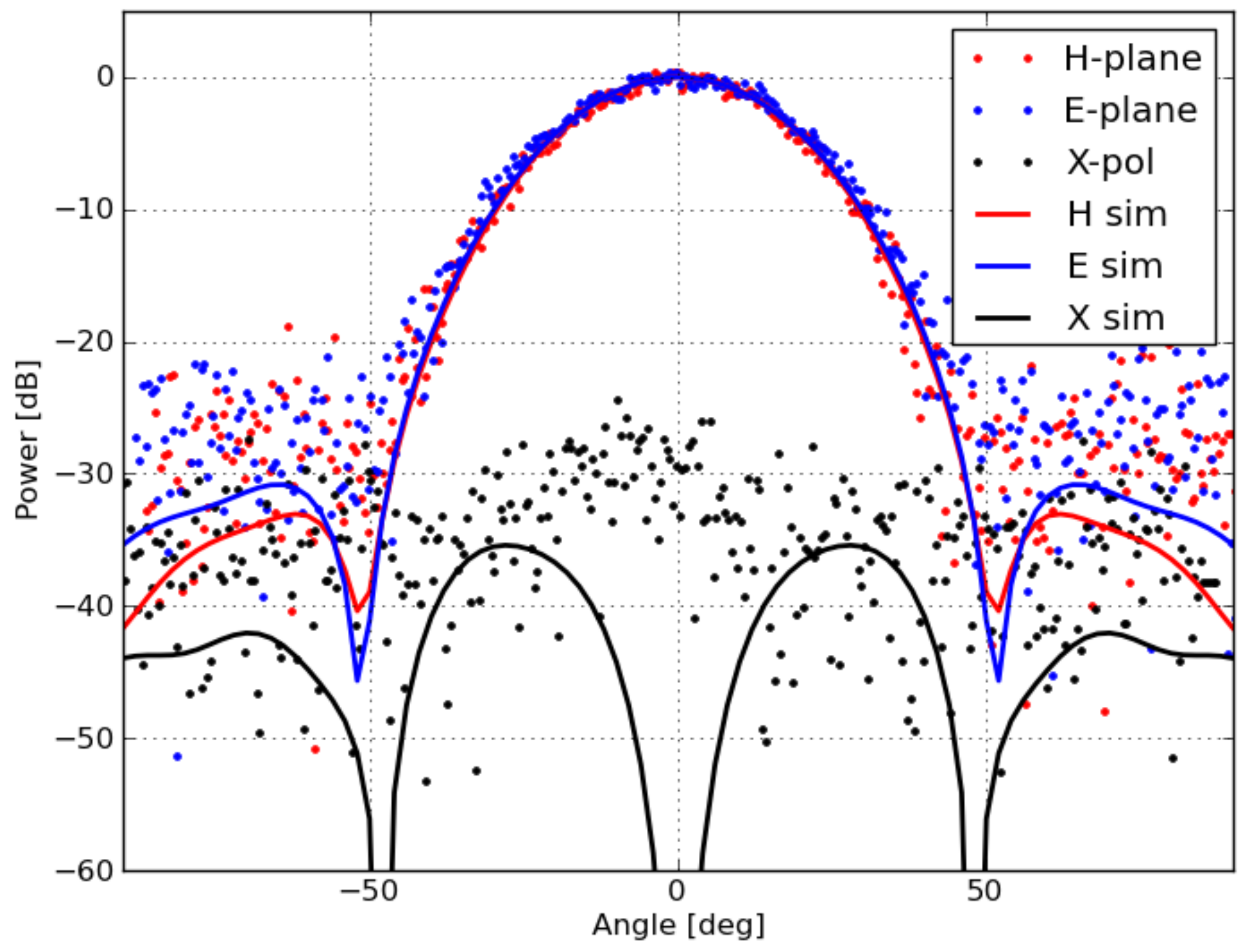}}
\caption{(Left) Return loss for six horns in one SPTpol feedhorn array.  Dashed line is the expectation for the feedhorn profile only.  (Middle) 300 K insertion loss measurements of the same horns.  (Right) Beam profile at 150 GHz for one representative horn.  Dots are measurements while solid lines are expectations from simulations.}\label{VNA}
\end{center}
\end{figure}

Before integration into 150 GHz modules, we measured the return loss and insertion loss for each silicon feedhorn array at 300 K.  Representative results for one feedhorn array are provided in Figure \ref{VNA}, where measurements for six separate horns are overplotted.  All horn arrays show return loss of $< -20$ dB at 300 K except for the frequency range 133-138 GHz.  The dashed line in the return loss plot represents the expectation for the feedhorn profile only, not including the square to circular waveguide section, which is known to have a return loss of $\sim -20$ dB.  The insertion loss at 300 K is in the middle plot of Figure \ref{VNA} and is $\sim -0.2\,\mbox{dB} \simeq 5\%$ on average.

\begin{figure}[h]
\begin{center}
\resizebox{0.8\textwidth}{!}{
\includegraphics[width=1.075\textwidth]{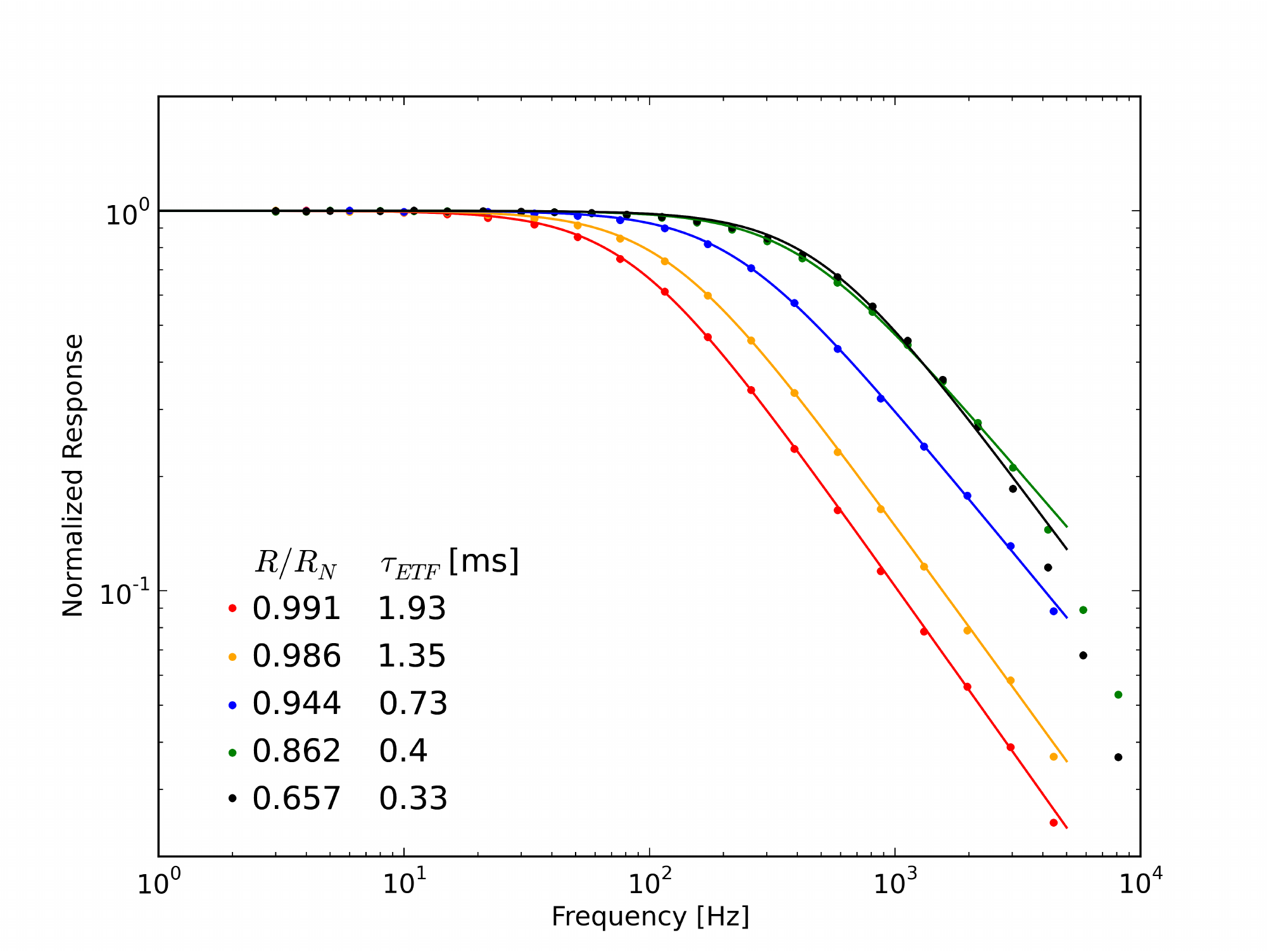}
\includegraphics[width=1.0\textwidth]{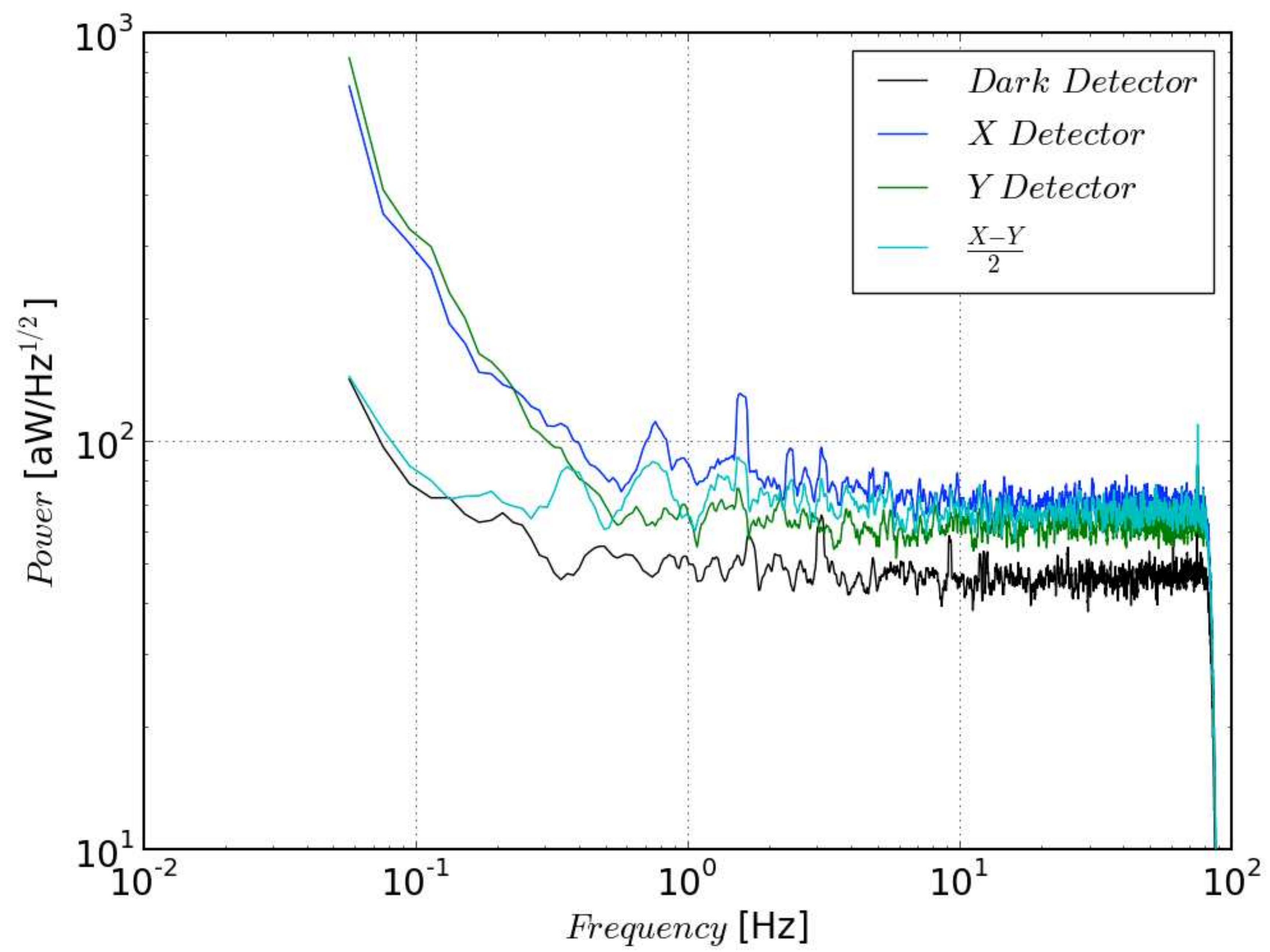}}
\caption{(Left) Measurements of the electrothermal time constant of one representative detector at several points in its superconducting transition.  Devices exhibit time constants $< 1$ ms at nominal operation points. (Right) Power spectral densities (PSDs) for three devices in the same 150~GHz pixel.  ``X" and ``Y" are optically loaded while the ``Dark" device is not coupled to the sky.  All three device PSDs have white noise levels consistent with expectations.  The PSD of the differenced timestreams of the optically loaded devices shows a much reduced $1/f$ knee.}\label{tau}
\end{center}
\end{figure}

Figure \ref{VNA} (Right) shows pre-deployment 150~GHz VNA measurements of one representative feedhorn.  Dots are measurements of the H-plane, E-plane, and cross-polarization, while solid lines are expectations from simulations.  The beam power drops below -20 dB at $\sim \pm 40^{\circ}$ from the beam center, and cross-polarization power is below -25 dB.  Measured beams of the detectors as deployed on the telescope also appear nominal, with an average full width half maximum (FWHM) of 1.06 arcminutes and beam eccentricity of $e=0.04$ for the 150 GHz pixels\cite{George_2012}.

Optical efficiency measurements were taken for a small subset of detectors prior to SPTpol deployment.  Using a set of metal mesh capacitive low-pass filters to define the upper edge of our bandpass, we illuminate the detectors with radiation from a cold load set to several temperatures between 4 and 30 K.  The measured difference in power compared to the expected in-band power gives the detector plus feedhorn optical efficiency\cite{Henning_2010}.  These measurements indicated detector optical efficiencies of $\sim 90\%$.

We measured the electrothermal time constants of many devices in the lab at various points in their superconducting transitions.  We apply an AC voltage bias of frequency $\omega$ to a detector just as we would in standard operation, but also apply additional voltage bias at a second bias tone $\omega + \delta\omega$, which amplitude-modulates the detector response.  The output current amplitude at the negative sideband at $\omega - \delta\omega$ (where no external voltage is applied) is a pure measurement of the electrothermal response of the device\cite{Lueker_2009}.  We plot the response as a function of $\delta\omega$ and fit a single pole to the data, which gives us the time constant of the device.  A representative plot of electrothermal time constant measurements is shown in Figure \ref{tau} (Left).  These measurements are taken several times when the device is biased at different points in its transition.  Across the arrays the electrothermal time constants are generally $< 1$ ms while in the superconducting transition.

We have also measured detector noise between field observations while on the telescope.  Figure \ref{tau} (Right) shows the noise for three detectors in a single representative 150~GHz pixel.  Detectors ``X" and ``Y" are optically coupled and looking at the sky, while the ``Dark" detector is not optically active.  Given in-lab calibration factors with systematics at the 10 - 20\% level, the white noise levels of all the devices are consistent with expectations, 47 aW/$\sqrt{\rm{Hz}}$ with no optical loading and 76 aW/$\sqrt{\rm{Hz}}$ with nominal optical load.  Additionally, differencing optically loaded detector timestreams removes correlated long time scale atmospheric fluctuations.  As a result, the power spectral density of the differenced timestreams shows a significant reduction in the $1/f$ knee, increasing the frequency range that can be used for extracting relevant science.

\section{Conclusions}\label{Conclusions}
SPTpol is a dichroic polarization-sensitive receiver recently deployed at the South Pole observing at 90 and 150 GHz.  We have discussed the design and properties of seven independent 150 GHz camera modules produced for the SPTpol focal plane.  The modules each contain an 84-pixel detector array (168 optical detectors) and a platelet array of single-moded corrugated feedhorns.  Mounting hardware absorbs differences in thermal contraction between module components to protect the silicon detector and feedhorn arrays from shattering.  The modules are sealed to RF radiation and contain a set of individual passive readout electronics.  Each module, therefore, is a self-contained camera that is easily tested and installed independently.

We measured detector properties of devices in five of the seven deployed arrays.  The arrays exhibit good uniformity in detector critical temperature $T_c$, normal resistance $R_N$, and saturation power $P_{sat}$, and the averages for all of these values across all tested devices are $T_c = 478.0 \pm 28.6$ mK , $R_N = 1.2\pm 0.2\,\Omega$, and $P_{sat} = 22.5\pm 5.7$ pW.  Furthermore, the detectors have white noise levels consistent with expectations.  We also obtained optical efficiencies for a small subset of 150 GHz devices, indicating $\sim 90\%$ detector efficiency given an assumed operating bandpass.  Detector electrothermal time constants are less than 1 ms at typical operation points, and preliminary beam measurements show a FWHM of 1.06 arcminutes and eccentricity $e=0.04$.

\acknowledgments   
Work at the University of Colorado - Boulder, Case Western Reserve University, the University of California - Berkeley, and the University of Chicago is supported by grants from the NSF (awards ANT-0638937, AST-0956135, and PHY-0114422), the Kavli Foundation, and the Gordon and Betty Moore Foundation.  Work at NIST is supported by the NIST Innovations in Measurement Science program.  The McGill authors acknowledge funding from the Natural Sciences and Engineering Research Council, Canadian Institute for Advanced Research, and Canada Research Chairs program. MD acknowledges support from an Alfred P. Sloan Research Fellowship.  Work at Argonne National Lab is supported by UChicago Argonne, LLC, Operator of Argonne National Laboratory (``Argonne").  Argonne, a U.S. Department of Energy Office of Science Laboratory, is operated under Contract No. DE-AC02-06CH11357.

\bibliographystyle{spiebib}   
\bibliography{spie_2012_henning}   

\begin{thebibliography}{10}

\bibitem{Chiang_2010}
H.~C. {Chiang}, P.~A.~R. {Ade}, D.~{Barkats}, J.~O. {Battle}, E.~M. {Bierman},
  J.~J. {Bock}, C.~D. {Dowell}, L.~{Duband}, E.~F. {Hivon}, W.~L. {Holzapfel},
  V.~V. {Hristov}, W.~C. {Jones}, B.~G. {Keating}, J.~M. {Kovac}, C.~L. {Kuo},
  A.~E. {Lange}, E.~M. {Leitch}, P.~V. {Mason}, T.~{Matsumura}, H.~T. {Nguyen},
  N.~{Ponthieu}, C.~{Pryke}, S.~{Richter}, G.~{Rocha}, C.~{Sheehy}, Y.~D.
  {Takahashi}, J.~E. {Tolan}, and K.~W. {Yoon}, ``{Measurement of Cosmic
  Microwave Background Polarization Power Spectra from Two Years of BICEP
  Data},'' ~{\bf 711}, pp.~1123--1140, Mar. 2010.

\bibitem{Larson_2011}
D.~{Larson}, J.~{Dunkley}, G.~{Hinshaw}, E.~{Komatsu}, M.~R. {Nolta}, C.~L.
  {Bennett}, B.~{Gold}, M.~{Halpern}, R.~S. {Hill}, N.~{Jarosik}, A.~{Kogut},
  M.~{Limon}, S.~S. {Meyer}, N.~{Odegard}, L.~{Page}, K.~M. {Smith}, D.~N.
  {Spergel}, G.~S. {Tucker}, J.~L. {Weiland}, E.~{Wollack}, and E.~L. {Wright},
  ``{Seven-year Wilkinson Microwave Anisotropy Probe (WMAP) Observations: Power
  Spectra and WMAP-derived Parameters},'' ~{\bf 192}, p.~16, Feb. 2011.

\bibitem{Lueker_2010}
M.~{Lueker}, C.~L. {Reichardt}, K.~K. {Schaffer}, O.~{Zahn}, P.~A.~R. {Ade},
  K.~A. {Aird}, B.~A. {Benson}, L.~E. {Bleem}, J.~E. {Carlstrom}, C.~L.
  {Chang}, H.-M. {Cho}, T.~M. {Crawford}, A.~T. {Crites}, T.~{de Haan}, M.~A.
  {Dobbs}, E.~M. {George}, N.~R. {Hall}, N.~W. {Halverson}, G.~P. {Holder},
  W.~L. {Holzapfel}, J.~D. {Hrubes}, M.~{Joy}, R.~{Keisler}, L.~{Knox}, A.~T.
  {Lee}, E.~M. {Leitch}, J.~J. {McMahon}, J.~{Mehl}, S.~S. {Meyer}, J.~J.
  {Mohr}, T.~E. {Montroy}, S.~{Padin}, T.~{Plagge}, C.~{Pryke}, J.~E. {Ruhl},
  L.~{Shaw}, E.~{Shirokoff}, H.~G. {Spieler}, B.~{Stalder}, Z.~{Staniszewski},
  A.~A. {Stark}, K.~{Vanderlinde}, J.~D. {Vieira}, and R.~{Williamson},
  ``{Measurements of Secondary Cosmic Microwave Background Anisotropies with
  the South Pole Telescope},'' ~{\bf 719}, pp.~1045--1066, Aug. 2010.

\bibitem{Das_2011}
S.~{Das}, T.~A. {Marriage}, P.~A.~R. {Ade}, P.~{Aguirre}, M.~{Amiri}, J.~W.
  {Appel}, L.~F. {Barrientos}, E.~S. {Battistelli}, J.~R. {Bond}, B.~{Brown},
  B.~{Burger}, J.~{Chervenak}, M.~J. {Devlin}, S.~R. {Dicker}, W.~{Bertrand
  Doriese}, J.~{Dunkley}, R.~{D{\"u}nner}, T.~{Essinger-Hileman}, R.~P.
  {Fisher}, J.~W. {Fowler}, A.~{Hajian}, M.~{Halpern}, M.~{Hasselfield},
  C.~{Hern{\'a}ndez-Monteagudo}, G.~C. {Hilton}, M.~{Hilton}, A.~D. {Hincks},
  R.~{Hlozek}, K.~M. {Huffenberger}, D.~H. {Hughes}, J.~P. {Hughes},
  L.~{Infante}, K.~D. {Irwin}, J.~{Baptiste Juin}, M.~{Kaul}, J.~{Klein},
  A.~{Kosowsky}, J.~M. {Lau}, M.~{Limon}, Y.-T. {Lin}, R.~H. {Lupton},
  D.~{Marsden}, K.~{Martocci}, P.~{Mauskopf}, F.~{Menanteau}, K.~{Moodley},
  H.~{Moseley}, C.~B. {Netterfield}, M.~D. {Niemack}, M.~R. {Nolta}, L.~A.
  {Page}, L.~{Parker}, B.~{Partridge}, B.~{Reid}, N.~{Sehgal}, B.~D. {Sherwin},
  J.~{Sievers}, D.~N. {Spergel}, S.~T. {Staggs}, D.~S. {Swetz}, E.~R.
  {Switzer}, R.~{Thornton}, H.~{Trac}, C.~{Tucker}, R.~{Warne}, E.~{Wollack},
  and Y.~{Zhao}, ``{The Atacama Cosmology Telescope: A Measurement of the
  Cosmic Microwave Background Power Spectrum at 148 and 218 GHz from the 2008
  Southern Survey},'' ~{\bf 729}, p.~62, Mar. 2011.

\bibitem{Keisler_2011}
R.~{Keisler}, C.~L. {Reichardt}, K.~A. {Aird}, B.~A. {Benson}, L.~E. {Bleem},
  J.~E. {Carlstrom}, C.~L. {Chang}, H.~M. {Cho}, T.~M. {Crawford}, A.~T.
  {Crites}, T.~{de Haan}, M.~A. {Dobbs}, J.~{Dudley}, E.~M. {George}, N.~W.
  {Halverson}, G.~P. {Holder}, W.~L. {Holzapfel}, S.~{Hoover}, Z.~{Hou}, J.~D.
  {Hrubes}, M.~{Joy}, L.~{Knox}, A.~T. {Lee}, E.~M. {Leitch}, M.~{Lueker},
  D.~{Luong-Van}, J.~J. {McMahon}, J.~{Mehl}, S.~S. {Meyer}, M.~{Millea}, J.~J.
  {Mohr}, T.~E. {Montroy}, T.~{Natoli}, S.~{Padin}, T.~{Plagge}, C.~{Pryke},
  J.~E. {Ruhl}, K.~K. {Schaffer}, L.~{Shaw}, E.~{Shirokoff}, H.~G. {Spieler},
  Z.~{Staniszewski}, A.~A. {Stark}, K.~{Story}, A.~{van Engelen},
  K.~{Vanderlinde}, J.~D. {Vieira}, R.~{Williamson}, and O.~{Zahn}, ``{A
  Measurement of the Damping Tail of the Cosmic Microwave Background Power
  Spectrum with the South Pole Telescope},'' ~{\bf 743}, p.~28, Dec. 2011.

\bibitem{Kovac_2002}
J.~M. {Kovac}, E.~M. {Leitch}, C.~{Pryke}, J.~E. {Carlstrom}, N.~W.
  {Halverson}, and W.~L. {Holzapfel}, ``{Detection of polarization in the
  cosmic microwave background using DASI},'' ~{\bf 420}, pp.~772--787, Dec.
  2002.

\bibitem{Barkats_2005b}
D.~{Barkats}, C.~{Bischoff}, P.~{Farese}, L.~{Fitzpatrick}, T.~{Gaier}, J.~O.
  {Gundersen}, M.~M. {Hedman}, L.~{Hyatt}, J.~J. {McMahon}, D.~{Samtleben},
  S.~T. {Staggs}, K.~{Vanderlinde}, and B.~{Winstein}, ``{First Measurements of
  the Polarization of the Cosmic Microwave Background Radiation at Small
  Angular Scales from CAPMAP},'' ~{\bf 619}, pp.~L127--L130, Feb. 2005.

\bibitem{Montroy_2006}
T.~E. {Montroy}, P.~A.~R. {Ade}, J.~J. {Bock}, J.~R. {Bond}, J.~{Borrill},
  A.~{Boscaleri}, P.~{Cabella}, C.~R. {Contaldi}, B.~P. {Crill}, P.~{de
  Bernardis}, G.~{De Gasperis}, A.~{de Oliveira-Costa}, G.~{De Troia}, G.~{di
  Stefano}, E.~{Hivon}, A.~H. {Jaffe}, T.~S. {Kisner}, W.~C. {Jones}, A.~E.
  {Lange}, S.~{Masi}, P.~D. {Mauskopf}, C.~J. {MacTavish}, A.~{Melchiorri},
  P.~{Natoli}, C.~B. {Netterfield}, E.~{Pascale}, F.~{Piacentini},
  D.~{Pogosyan}, G.~{Polenta}, S.~{Prunet}, S.~{Ricciardi}, G.~{Romeo}, J.~E.
  {Ruhl}, P.~{Santini}, M.~{Tegmark}, M.~{Veneziani}, and N.~{Vittorio}, ``{A
  Measurement of the CMB EE Spectrum from the 2003 Flight of BOOMERANG},''
  ~{\bf 647}, pp.~813--822, 2006.

\bibitem{Bischoff_2008}
C.~{Bischoff}, L.~{Hyatt}, J.~J. {McMahon}, G.~W. {Nixon}, D.~{Samtleben},
  K.~M. {Smith}, K.~{Vanderlinde}, D.~{Barkats}, P.~{Farese}, T.~{Gaier}, J.~O.
  {Gundersen}, M.~M. {Hedman}, S.~T. {Staggs}, and B.~{Winstein}, ``{New
  Measurements of Fine-Scale CMB Polarization Power Spectra from CAPMAP at Both
  40 and 90 GHz},'' ~{\bf 684}, pp.~771--789, Sept. 2008.

\bibitem{Pryke_2009}
C.~{Pryke}, P.~{Ade}, J.~{Bock}, M.~{Bowden}, M.~L. {Brown}, G.~{Cahill}, P.~G.
  {Castro}, S.~{Church}, T.~{Culverhouse}, R.~{Friedman}, K.~{Ganga}, W.~K.
  {Gear}, S.~{Gupta}, J.~{Hinderks}, J.~{Kovac}, A.~E. {Lange}, E.~{Leitch},
  S.~J. {Melhuish}, Y.~{Memari}, J.~A. {Murphy}, A.~{Orlando}, R.~{Schwarz},
  C.~O. {Sullivan}, L.~{Piccirillo}, N.~{Rajguru}, B.~{Rusholme}, A.~N.
  {Taylor}, K.~L. {Thompson}, A.~H. {Turner}, E.~Y.~S. {Wu}, and M.~{Zemcov},
  ``{Second and Third Season QUaD Cosmic Microwave Background Temperature and
  Polarization Power Spectra},'' ~{\bf 692}, pp.~1247--1270, Feb. 2009.

\bibitem{Bleem_2012}
L.~{Bleem}, P.~{Ade}, K.~{Aird}, J.~{Austermann}, J.~{Beall}, D.~{Becker},
  B.~{Benson}, J.~{Britton}, J.~{Carlstrom}, C.~L. {Chang}, H.~{Cho}, T.~{de
  Haan}, T.~{Crawford}, A.~{Crites}, A.~{Datesman}, M.~{Dobbs}, W.~{Everett},
  A.~{Ewall-Wice}, E.~{George}, N.~{Halverson}, N.~{Harrington}, J.~{Henning},
  G.~{Hilton}, W.~{Holzapfel}, S.~{Hoover}, J.~{Hubmayr}, K.~{Irwin},
  R.~{Keisler}, J.~{Kennedy}, A.~{Lee}, E.~{Leitch}, D.~{Li}, M.~{Lueker},
  D.~P. {Marrone}, J.~{McMahon}, J.~{Mehl}, S.~{Meyer}, J.~{Montgomery},
  T.~{Montroy}, T.~{Natoli}, J.~{Nibarger}, M.~{Niemack}, V.~{Novosad},
  S.~{Padin}, C.~{Pryke}, C.~{Reichardt}, J.~{Ruhl}, B.~{Saliwanchik},
  J.~{Sayre}, K.~{Schafer}, E.~{Shirokoff}, K.~{Story}, K.~{Vanderlinde},
  J.~{Vieira}, G.~{Wang}, R.~{Williamson}, V.~{Yefremenko}, K.~W. {Yoon}, and
  E.~{Young}, ``{An Overview of the SPTpol Experiment},'' {\em Journal of Low
  Temperature Physics}~{\bf 167}, pp.~859--864, June 2012.

\bibitem{Austermann_2012}
J.~{Austermann}, K.~{Aird}, J.~{Beall}, D.~{Becker}, A.~{Bender}, B.~{Benson},
  L.~{Bleem}, J.~{Britton}, J.~{Carlstrom}, C.~{Chang}, H.~{Chiang}, H.-M.
  {Cho}, T.~{Crawford}, A.~{Crites}, A.~{Datesman}, T.~{de Haan}, M.~{Dobbs},
  E.~{George}, N.~{Halverson}, N.~{Harrington}, J.~{Henning}, G.~{Hilton},
  G.~{Holder}, W.~{Holzapfel}, S.~{Hoover}, N.~{Huang}, J.~{Hubmayr},
  K.~{Irwin}, R.~{Keisler}, J.~{Kennedy}, L.~{Knox}, A.~{Lee}, E.~{Leitch},
  D.~{Li}, M.~{Lueker}, D.~{Marrone}, J.~{McMahon}, J.~{Mehl}, S.~{Meyer},
  T.~{Montroy}, T.~{Natoli}, J.~{Nibarger}, M.~{Niemack}, V.~{Novosad},
  S.~{Padin}, C.~{Pryke}, C.~{Reichardt}, J.~{Ruhl}, B.~{Saliwanchik},
  J.~{Sayre}, K.~{Schaffer}, E.~{Shirokoff}, A.~{Stark}, K.~{Story},
  K.~{Vanderlinde}, J.~{Vieira}, G.~{Wang}, R.~{Williamson}, V.~{Yefremenko},
  K.~W. {Yoon}, and O.~{Zahn}, ``{An instrument for CMB polarization
  measurements with the South Pole Telescope},'' in {\em Society of
  Photo-Optical Instrumentation Engineers (SPIE) Conference Series},  July
  2012.

\bibitem{Sayre_2012}
J.~{Sayre}, P.~{Ade}, K.~{Aird}, J.~{Austermann}, J.~{Beall}, D.~{Becker},
  B.~{Benson}, L.~{Bleem}, J.~{Britton}, J.~{Carlstrom}, C.~{Chang}, H.-M.
  {Cho}, T.~{Crawford}, A.~{Crites}, A.~{Datesman}, T.~{de Haan}, M.~{Dobbs},
  W.~{Everett}, A.~{Ewall-Wice}, E.~{George}, N.~{Halverson}, N.~{Harrington},
  J.~{Henning}, G.~{Hilton}, W.~{Holzapfel}, J.~{Hubmayr}, K.~{Irwin},
  M.~{Karfunkle}, R.~{Keisler}, J.~{Kennedy}, A.~{Lee}, E.~{Leitch}, D.~{Li},
  M.~{Lueker}, D.~{Marrone}, J.~{McMahon}, J.~{Mehl}, S.~{Meyer},
  J.~{Montgomery}, T.~{Montroy}, T.~{Natoli}, J.~{Nibarger}, M.~{Niemack},
  V.~{Novosad}, S.~{Padin}, C.~{Pryke}, C.~{Reichardt}, J.~{Ruhl},
  B.~{Saliwanchik}, J.~{Sayre}, K.~{Schaffer}, E.~{Shirokoff}, K.~{Story},
  C.~{Tucker}, K.~{Vanderlinde}, J.~{Vieira}, G.~{Wang}, R.~{Williamson},
  V.~{Yefremenko}, K.~W. {Yoon}, and E.~{Young}, ``{Design and characterization
  of 90 GHz feedhorn-coupled TES polarimeter pixels in the SPTPol camera},'' in
  {\em Society of Photo-Optical Instrumentation Engineers (SPIE) Conference
  Series},  July 2012.

\bibitem{Clarricoats_1984}
P.~J.~B. Clarricoats and A.~D. Olver, {\em Corrugated horns for microwave
  antennas / P.J.B. Clarricoats and A.D. Olver}, P. Peregrinus on behalf of the
  Institution of Electrical Engineers, London, UK :, 1984.

\bibitem{Barnes_2002}
C.~{Barnes}, M.~{Limon}, L.~{Page}, C.~{Bennett}, S.~{Bradley}, M.~{Halpern},
  G.~{Hinshaw}, N.~{Jarosik}, W.~{Jones}, A.~{Kogut}, S.~{Meyer},
  O.~{Motrunich}, G.~{Tucker}, D.~{Wilkinson}, and E.~{Wollack}, ``{The MAP
  Satellite Feed Horns},'' ~{\bf 143}, pp.~567--576, Dec. 2002.

\bibitem{Padin_2002}
S.~{Padin}, M.~C. {Shepherd}, J.~K. {Cartwright}, R.~G. {Keeney}, B.~S.
  {Mason}, T.~J. {Pearson}, A.~C.~S. {Readhead}, W.~A. {Schaal}, J.~{Sievers},
  P.~S. {Udomprasert}, J.~K. {Yamasaki}, W.~L. {Holzapfel}, J.~E. {Carlstrom},
  M.~{Joy}, S.~T. {Myers}, and A.~{Otarola}, ``{The Cosmic Background
  Imager},'' ~{\bf 114}, pp.~83--97, Jan. 2002.

\bibitem{Jones_2003}
W.~C. {Jones}, R.~{Bhatia}, J.~J. {Bock}, and A.~E. {Lange}, ``{A Polarization
  Sensitive Bolometric Receiver for Observations of the Cosmic Microwave
  Background},'' in {\em Society of Photo-Optical Instrumentation Engineers
  (SPIE) Conference Series},  T.~G. {Phillips} and J.~{Zmuidzinas}, eds., {\em
  Society of Photo-Optical Instrumentation Engineers (SPIE) Conference Series}
  {\bf 4855}, pp.~227--238, Feb. 2003.

\bibitem{Barkats_2005}
D.~{Barkats}, C.~{Bischoff}, P.~{Farese}, T.~{Gaier}, J.~O. {Gundersen}, M.~M.
  {Hedman}, L.~{Hyatt}, J.~J. {McMahon}, D.~{Samtleben}, S.~T. {Staggs},
  E.~{Stefanescu}, K.~{Vanderlinde}, and B.~{Winstein}, ``{Cosmic Microwave
  Background Polarimetry Using Correlation Receivers with the PIQUE and CAPMAP
  Experiments},'' ~{\bf 159}, pp.~1--26, July 2005.

\bibitem{Bock_2009}
J.~J. {Bock}, J.~{Gundersen}, A.~T. {Lee}, P.~L. {Richards}, and E.~{Wollack},
  ``{Optical coupling},'' {\em Journal of Physics Conference Series}~{\bf 155},
  p.~012005, Mar. 2009.

\bibitem{Hinderks_2009}
J.~R. {Hinderks}, P.~{Ade}, J.~{Bock}, M.~{Bowden}, M.~L. {Brown}, G.~{Cahill},
  J.~E. {Carlstrom}, P.~G. {Castro}, S.~{Church}, T.~{Culverhouse},
  R.~{Friedman}, K.~{Ganga}, W.~K. {Gear}, S.~{Gupta}, J.~{Harris},
  V.~{Haynes}, B.~G. {Keating}, J.~{Kovac}, E.~{Kirby}, A.~E. {Lange},
  E.~{Leitch}, O.~E. {Mallie}, S.~{Melhuish}, Y.~{Memari}, A.~{Murphy},
  A.~{Orlando}, R.~{Schwarz}, C.~O. {Sullivan}, L.~{Piccirillo}, C.~{Pryke},
  N.~{Rajguru}, B.~{Rusholme}, A.~N. {Taylor}, K.~L. {Thompson}, C.~{Tucker},
  A.~H. {Turner}, E.~Y.~S. {Wu}, and M.~{Zemcov}, ``{QUaD: A High-Resolution
  Cosmic Microwave Background Polarimeter},'' ~{\bf 692}, pp.~1221--1246, Feb.
  2009.

\bibitem{Takahashi_2010}
Y.~D. {Takahashi}, P.~A.~R. {Ade}, D.~{Barkats}, J.~O. {Battle}, E.~M.
  {Bierman}, J.~J. {Bock}, H.~C. {Chiang}, C.~D. {Dowell}, L.~{Duband}, E.~F.
  {Hivon}, W.~L. {Holzapfel}, V.~V. {Hristov}, W.~C. {Jones}, B.~G. {Keating},
  J.~M. {Kovac}, C.~L. {Kuo}, A.~E. {Lange}, E.~M. {Leitch}, P.~V. {Mason},
  T.~{Matsumura}, H.~T. {Nguyen}, N.~{Ponthieu}, C.~{Pryke}, S.~{Richter},
  G.~{Rocha}, and K.~W. {Yoon}, ``{Characterization of the BICEP Telescope for
  High-precision Cosmic Microwave Background Polarimetry},'' ~{\bf 711},
  pp.~1141--1156, Mar. 2010.

\bibitem{Britton_2010}
J.~W. {Britton}, J.~P. {Nibarger}, K.~W. {Yoon}, J.~A. {Beall}, D.~{Becker},
  H.-M. {Cho}, G.~C. {Hilton}, J.~{Hubmayr}, M.~D. {Niemack}, and K.~D.
  {Irwin}, ``{Corrugated silicon platelet feed horn array for CMB polarimetry
  at 150 GHz},'' in {\em Society of Photo-Optical Instrumentation Engineers
  (SPIE) Conference Series},  {\em Society of Photo-Optical Instrumentation
  Engineers (SPIE) Conference Series} {\bf 7741}, July 2010.

\bibitem{Hubmayr_2012a}
J.~{Hubmayr}, J.~W. {Appel}, J.~E. {Austermann}, J.~A. {Beall}, D.~{Becker},
  B.~A. {Benson}, L.~E. {Bleem}, J.~E. {Carlstrom}, C.~L. {Chang}, H.~M. {Cho},
  A.~T. {Crites}, T.~{Essinger-Hileman}, A.~{Fox}, E.~M. {George}, N.~W.
  {Halverson}, N.~L. {Harrington}, J.~W. {Henning}, G.~C. {Hilton}, W.~L.
  {Holzapfel}, K.~D. {Irwin}, A.~T. {Lee}, D.~{Li}, J.~{McMahon}, J.~{Mehl},
  T.~{Natoli}, M.~D. {Niemack}, L.~B. {Newburgh}, J.~P. {Nibarger}, L.~P.
  {Parker}, B.~L. {Schmitt}, S.~T. {Staggs}, J.~{Van Lanen}, E.~J. {Wollack},
  and K.~W. {Yoon}, ``{An All Silicon Feedhorn-Coupled Focal Plane for Cosmic
  Microwave Background Polarimetry},'' {\em Journal of Low Temperature
  Physics}~{\bf 167}, pp.~904--910, June 2012.

\bibitem{Austermann_2009}
J.~E. {Austermann}, M.~D. {Niemack}, J.~W. {Appel}, J.~A. {Beall}, D.~{Becker},
  D.~A. {Bennett}, B.~A. {Benson}, L.~E. {Bleem}, J.~{Britton}, J.~E.
  {Carlstrom}, C.~L. {Chang}, H.~M. {Cho}, A.~T. {Crites},
  T.~{Essinger-Hileman}, W.~{Everett}, N.~W. {Halverson}, J.~W. {Henning},
  G.~C. {Hilton}, K.~D. {Irwin}, J.~{McMahon}, J.~{Mehl}, S.~S. {Meyer}, L.~P.
  {Parker}, S.~M. {Simon}, S.~T. {Staggs}, J.~N. {Ullom}, C.~{Visnjic}, K.~W.
  {Yoon}, and Y.~{Zhao}, ``{Measurements of Bolometer Uniformity for Feedhorn
  Coupled TES Polarimeters},'' in {\em American Institute of Physics Conference
  Series},  {B.~Young, B.~Cabrera, \& A.~Miller}, ed., {\em American Institute
  of Physics Conference Series} {\bf 1185}, pp.~498--501, Dec. 2009.

\bibitem{Bleem_2009}
L.~E. {Bleem}, J.~W. {Appel}, J.~E. {Austermann}, J.~A. {Beall}, D.~T.
  {Becker}, B.~A. {Benson}, J.~{Britton}, J.~E. {Carlstrom}, C.~L. {Chang},
  H.~M. {Cho}, A.~T. {Crites}, T.~{Essinger-Hileman}, W.~{Everett}, N.~W.
  {Halverson}, J.~W. {Henning}, G.~C. {Hilton}, K.~D. {Irwin}, J.~{McMahon},
  J.~{Mehl}, S.~S. {Meyer}, M.~D. {Niemack}, L.~P. {Parker}, S.~M. {Simon},
  S.~T. {Staggs}, C.~{Visnjic}, K.~W. {Yoon}, and Y.~{Zhao}, ``{Optical
  properties of Feedhorn-coupled TES polarimeters for CMB polarimetry},'' in
  {\em American Institute of Physics Conference Series},  {B.~Young,
  B.~Cabrera, \& A.~Miller}, ed., {\em American Institute of Physics Conference
  Series} {\bf 1185}, pp.~479--482, Dec. 2009.

\bibitem{Yoon_2009}
K.~W. {Yoon}, J.~W. {Appel}, J.~E. {Austermann}, J.~A. {Beall}, D.~{Becker},
  B.~A. {Benson}, L.~E. {Bleem}, J.~{Britton}, C.~L. {Chang}, J.~E.
  {Carlstrom}, H.~{Cho}, A.~T. {Crites}, T.~{Essinger-Hileman}, W.~{Everett},
  N.~W. {Halverson}, J.~W. {Henning}, G.~C. {Hilton}, K.~D. {Irwin},
  J.~{McMahon}, J.~{Mehl}, S.~S. {Meyer}, S.~{Moseley}, M.~D. {Niemack}, L.~P.
  {Parker}, S.~M. {Simon}, S.~T. {Staggs}, K.~{U-Yen}, C.~{Visnjic},
  E.~{Wollack}, and Y.~{Zhao}, ``{Feedhorn-Coupled TES Polarimeters for
  Next-Generation CMB Instruments},'' in {\em American Institute of Physics
  Conference Series},  {B.~Young, B.~Cabrera, \& A.~Miller}, ed., {\em American
  Institute of Physics Conference Series} {\bf 1185}, pp.~515--518, Dec. 2009.

\bibitem{Henning_2010}
J.~W. {Henning}, J.~W. {Appel}, J.~E. {Austermann}, J.~A. {Beall}, D.~{Becker},
  D.~A. {Bennett}, L.~E. {Bleem}, B.~A. {Benson}, J.~{Britton}, J.~E.
  {Carlstrom}, C.~L. {Chang}, H.~M. {Cho}, A.~T. {Crites},
  T.~{Essinger-Hileman}, W.~{Everett}, E.~M. {George}, N.~W. {Halverson}, G.~C.
  {Hilton}, W.~L. {Holzapfel}, J.~{Hubmayr}, K.~D. {Irwin}, D.~{Li},
  J.~{McMahon}, J.~{Mehl}, S.~S. {Meyer}, S.~{Moseley}, J.~P. {Nibarger}, M.~D.
  {Niemack}, L.~P. {Parker}, E.~{Shirokoff}, S.~M. {Simon}, S.~T. {Staggs},
  J.~N. {Ullom}, K.~{U-Yen}, C.~{Visnjic}, E.~{Wollack}, K.~W. {Yoon}, E.~Y.
  {Young}, and Y.~{Zhao}, ``{Optical efficiency of feedhorn-coupled TES
  polarimeters for next-generation CMB instruments},'' in {\em Society of
  Photo-Optical Instrumentation Engineers (SPIE) Conference Series},  {\em
  Society of Photo-Optical Instrumentation Engineers (SPIE) Conference Series}
  {\bf 7741}, July 2010.

\bibitem{Smecher_2012}
G.~{Smecher}, F.~{Aubin}, E.~{Bissonnette}, M.~{Dobbs}, P.~{Hyland}, and
  K.~{MacDermid}, ``{A Biasing and Demodulation System for Kilopixel TES
  Bolometer Arrays},'' in {\em IEEE Transactions on Instrumentation and
  Measurement},   {\bf 61}, Jan. 2012.

\bibitem{eccosorb}
{Emerson and Cuming Microwave Products, http://www.eccosorb.com}

\bibitem{stycast}
{http://lartpc-docdb.fnal.gov/0000/000059/001/stycas2850.pdf}

\bibitem{George_2012}
E.~{George}, P.~{Ade}, K.~{Aird}, J.~{Austermann}, J.~{Beall}, D.~{Becker},
  A.~{Bender}, B.~{Benson}, L.~{Bleem}, J.~{Britton}, J.~{Carlstrom},
  C.~{Chang}, H.~{Chiang}, H.-M. {Cho}, T.~{Crawford}, A.~{Crites},
  A.~{Datesman}, T.~{de Haan}, M.~{Dobbs}, W.~{Everett}, A.~{Ewall-Wice},
  N.~{Halverson}, N.~{Harrington}, J.~{Henning}, G.~{Hilton}, W.~{Holzapfel},
  S.~{Hoover}, N.~{Huang}, J.~{Hubmayr}, K.~{Irwin}, M.~{Karfunkle},
  R.~{Keisler}, J.~{Kennedy}, A.~{Lee}, E.~{Leitch}, D.~{Li}, M.~{Lueker},
  D.~{Marrone}, J.~{McMahon}, J.~{Mehl}, S.~{Meyer}, J.~{Montgomery},
  T.~{Montroy}, J.~{Nagy}, T.~{Natoli}, J.~{Nibarger}, M.~{Niemack},
  V.~{Novosad}, S.~{Padin}, C.~{Pryke}, C.~{Reichardt}, J.~{Ruhl},
  B.~{Saliwanchik}, J.~{Sayre}, K.~{Schaffer}, E.~{Shirokoff}, K.~{Story},
  C.~{Tucker}, K.~{Vanderlinde}, J.~{Vieira}, G.~{Wang}, R.~{Williamson},
  V.~{Yefremenko}, K.~W. {Yoon}, and E.~{Young}, ``{Performance and on-sky
  optical characterization of the SPTpol instrument},'' in {\em Society of
  Photo-Optical Instrumentation Engineers (SPIE) Conference Series},  July
  2012.

\bibitem{Lueker_2009}
M.~{Lueker}, B.~A. {Benson}, C.~L. {Chang}, H.-M. {Cho}, M.~{Dobbs}, W.~L.
  {Holzapfel}, T.~{Lanting}, A.~T. {Lee}, J.~{Mehl}, T.~{Plagge},
  E.~{Shirokoff}, H.~G. {Spieler}, and J.~D. {Vieira}, ``{Thermal Design and
  Characterization of Transition-Edge Sensor (TES) Bolometers for
  Frequency-Domain Multiplexing},'' {\em IEEE Transactions on Applied
  Superconductivity}~{\bf 19}, pp.~496--500, June 2009.

\end{thebibliography}

\end{document}